\documentclass[12pt]{article}

\usepackage{tikz}
\usetikzlibrary{decorations.pathreplacing}

\usepackage{array,float,epsf,latexsym,psfrag,amssymb,eurosym,
theorem,amsmath,verbatim,framed}

\usepackage{graphicx}
\usepackage{oldgerm}
\usepackage{amsfonts}
\usepackage{bbm,mathbbol}
\usepackage{mathrsfs} 

\usepackage{color}

\usepackage{bm}
\DeclareMathAlphabet{\mathsfit}{T1}{\sfdefault}{\mddefault}{\sldefault}
\SetMathAlphabet{\mathsfit}{bold}{T1}{\sfdefault}{\bfdefault}{\sldefault}
\def\pt #1{\bm{\mathsfit{#1}}}

\DeclareMathAlphabet{\mathbbmsl}{U}{bbm}{m}{sl}
\def\el #1{\mathbbmsl{#1}}

\def\pw #1{{\sf #1}}

\def\tz #1{\mbox{\boldmath $#1$}}

\def\fr   #1{         \mathfrak{#1} }

\def\gz #1{\mbox{\boldmath $\mit #1$}}

\def\und  {\quad\mbox{and }\quad}
\def\with {\quad\mbox{with}\quad}

\def\Grad {\nabla_{X    }}
\def\grad {\nabla_{x    }}

\def\cof {\mbox{cof}\,}

\def\rot {\mbox{curl}}

\def\div {\mbox{div}}

\def\Div {\mbox{Div}}

\def\jmp #1{[\![ #1 ]\!]}

\def\lto {\;\Longrightarrow\;}
\def\cl #1{{\cal #1}}

\def\d {\,\mbox{d}}
\def\D {\,\mbox{D}}

\def\be {\begin{equation}}
\def\ee {\end  {equation}}
\def\bea {\begin{eqnarray}}
\def\eea {\end  {eqnarray}}
\def\bean {\begin{eqnarray*}}
\def\eean {\end  {eqnarray*}}
\def\ben {$$}
\def\een {$$}

\newcolumntype{L}[1]{>{\raggedright\let\newline\\\arraybackslash\hspace{0pt}}m{#1}}
\newcolumntype{C}[1]{>{\centering\let\newline\\\arraybackslash\hspace{0pt}}m{#1}}
\newcolumntype{R}[1]{>{\raggedleft\let\newline\\\arraybackslash\hspace{0pt}}m{#1}}

\begin{document}

\pagestyle{plain}
\pagenumbering{roman}
\cleardoublepage
\pagenumbering{arabic}

\begin{center}
{\Large {\bf A Geometrically Exact Continuum Framework for\\[2mm] Light-Matter Interaction in Photo-Active Polymers\\[2mm]
I. Variational Setting
}}\\[3mm]
M Mehnert$^1$, W Oates$^2$, P Steinmann$^{1,3}$\\[2mm]
$^1$ Institute of Applied Mechanics, Friedrich-Alexander University Erlangen-Nuremberg,\\ 91054 Erlangen, Germany\\[1mm]
$^2$ Florida Center for Advanced Aero Propulsion (FCAAP),\\
Department of Mechanical Engineering,Florida A \& M and Florida State University,\\Tallahassee, Florida 32310, USA\\[1mm]
$^3$ Glasgow Computational Engineering Centre, University of Glasgow,\\ G12 8QQ Glasgow, United Kingdom\\[6mm]

{\bf Abstract}
\end{center}

Molecular photo-switches as, e.g., azobenzene molecules allow, when embedded into a polymeric matrix, for photo-active polymer compounds responding mechanically when exposed to light of certain wavelength. Photo-mechanics, i.e.\ light-matter interaction in photo-active polymers holds great promise for, e.g., remote and contact-free activation of photo-driven actuators. In a series of earlier contributions, Oates et al. developed a successful continuum formulation for the coupled electric, electronic and mechanical problem capturing azobenzene polymer compounds, thereby mainly focussing on geometrically linearized kinematics  \cite{bin2015unified, oates2017rate,roberts2017photomechanically}. Building on that formulation, we here explore the variational setting of a geometrically exact continuum framework based on Dirichlet's and Hamilton's principle as well as, noteworthy, Hamilton's equations. Thereby, when treating the dissipative case, we resort to incremental versions of the various variational problems via suited incorporation of a dissipation potential. In particular, the Hamiltonian setting of geometrically exact photo-mechanics is up to now largely under-explored even for the energetic case, arguably since the corresponding Lagrangian is degenerate in Dirac's sense. Moreover, in general, the Hamiltonian setting of dissipative dynamical systems is a matter of ongoing debate per se. In this contribution, by advocating a novel incremental version of the Hamiltonian setting exemplified for the dissipative case of photo-mechanics, we aim to also unify the variational approach to dissipative dynamical systems. Taken together, the variational setting of a geometrically exact continuum framework of photo-mechanics paves the way for forthcoming theoretical and numerical analyses.

\section{Introduction}

Photosensitive materials possess the ability to convert photonic energy into a mechanical material response, which  eliminates the necessity of electric wiring or circuits of conventional smart materials such as shape memory alloys, electroceramics or electro-active polymers \cite{kundys2015photostrictive}. The photo-mechanical coupling can originate from various, fundamentally different physical effects, depending on the specific material under investigation.  In electrically polar solids, the bulk deformation results from the converse piezoelectric effect in combination with the photovoltaic effect. While this combination of properties was discovered in single crystals of SbSI \cite{PhysRevLett.17.198, fridkin1967current}, more recently ferroelectric compounds such as BiFeO$_3$ \cite{PhysRevLett.100.227602, kreisel2012photoferroelectric} and PbTiO$_3$ in form of thin films \cite{PhysRevLett.108.087601} have come into focus. The group of polar and non-polar semiconductors show a similar behavior, as the material deformation also originates from the converse piezoelectric effect. However, in the case of polar semiconductors the necessary electric field is generated by light-induced changes in the free surface charges \cite{lagowski1972photomechanical,lagowski1974photomechanical}  whereas in the case of non-polar semiconductors such as Germanium, Silicon or Carbon nanotubes, an excess of electron hole pairs induces the electric stimulus \cite{figielski1961photostriction,gauster1967electronic,zhang1999elastic,PhysRevB.76.165437}. Another promising representative of photo-sensitive materials can be found in organic polymers, in which conformational changes of molecular switches, e.g. from rod (low energy state) to kinked (high energy state) shape, are triggered by light and, when embedded into a polymeric matrix, result in (potentially large) photo-induced deformation of the resulting (effective) compound material \cite{finkelmann2001new,zhao2009smart}. As an example, in azobenzene photo-switches, these conformational changes are a consequence of trans-cis (and likewise the reverse cis-trans) photo-isomerization depending on the wave-length of the incident light, typically in the $\sim350 - 500$ nm (UV to visible) range \cite{wang2012trans, jiang2006polymers, luo2009constitutive}. Clearly, photo-sensitive polymers promise fascinating applications, e.g., for  remote and contact-free activation of optical actuators \cite{uchino1997new,uchino20125}.\\

Recently, in a series of contributions Oates et al.  proposed a comprehensive phenomenological continuum formulation of photo-mechanics that is specifically tailored to capture the light-induced mechanical response of azobenzene polymers \cite{bin2015unified, oates2017rate,roberts2017photomechanically}. Therein modeling at
continuum length scales, while retaining the specifics of the underlying light-matter interactions in a homogenized sense, relies on the introduction of electronic order parameter fields in addition to the electric and mechanical fields common in the field of electro-mechanics \cite{Mehnert1}. Thus, along with the electronic degrees of freedom, an electronic (micro-force-type) balance equation arises in addition to the common electric and mechanical balances. Taken together, the formulation sub-divides into an electric, an electronic and a mechanical sub-problem. The formulation in \cite{bin2015unified, oates2017rate,roberts2017photomechanically} focuses mainly on geometrically linearized kinematics when treating azobenzene-polymer compounds that qualify as mechanically stiff. Interestingly, ongoing research in organic chemistry focuses on synthesis and characterization of a variety of alternative polymer compounds involving various types of molecular photo-switches, thereby also promising options for mechanically soft photo-active polymers \cite{irie2008photochromism,yu2014recent,iqbal2013photo}. Consequently, a geometrically exact continuum modeling framework is a necessary tool for the analytical and computational design and optimisation of future photo-mechanical devices.\\

Motivated by this state of affairs, we here pursue a rigorous geometrically nonlinear account on the phenomenological continuum modeling of photo-active polymers, thereby focusing  on the variational setting.
When considering light-matter interaction, the frequency (wave length) of electro-magnetic waves associated with the incident light is orders of magnitude higher (smaller) than that displayed by the mechanical response of matter at the continuum length scale. Thus, for the sake of simplicity, we here consider all electric quantities as time-averaged, indeed as quasi-static, and neglect any magnetic effects. Consequently, only the electric potential and electric Gauss law remain from the electro-magnetic degrees of freedom and the set of Maxwell equations describing electro-magneto-dynamics.\\

The electronic and mechanical solution fields may, however, display inertia effects, thus requiring their incorporation at the continuum length scale. Consequently, after exercising the quasi-static case within the realm of Dirichlet's principle as a preliminary, we treat the dynamic case of the coupled problem via Hamilton's principle and, noteworthy, via Hamilton's equations. The latter is somewhat sophisticated due to the Lagrangian being degenerate in the sense of Dirac's theory since no velocity of the electric potential is involved. Thus, Legendre transformation of the Lagrangian into the Hamiltonian involves Lagrange multipliers to enforce corresponding constraints on the associated momenta \cite{dirac1950generalized,dirac1958generalized}.\\

Importantly, light-matter interaction is associated with energetic losses, e.g.\, due to optical scattering and/or photochemical reactions, thus asking for consideration of dissipation. We will thus in particular demonstrate how to cope with the dissipative case by resorting to incremental variational settings. Noteworthy, especially Hamilton's equations based on an incremental total energy are a novelty also beyond photo-mechanics.\\

We structure this manuscript as follows: Section 2 first introduces the electric, electronic and mechanical solution fields and their space-time gradients in the context of a geometrically exact continuum description. Based thereon, Section 3 discusses the corresponding contributions to the various internal and external energy densities. These serve as potentials for the explicit constitutive relations in Section 4. To set the stage for the variational setting of photo-mechanics, Section 5 explores the energetic and dissipative case of Dirichlet's principle as a preliminary to a corresponding account on Hamilton's principle in Section 6. Lastly, Section 7 addresses the two-fold challenge associated with Hamilton's equations, i.e.\ the determination of the Hamiltonian from a degenerate Lagrangian and a proper account for the dissipative case. Eventually, Section 8 concludes the manuscript.\\

\section{Preliminaries}
The subsequently formulated continuum framework for the description of light-matter interaction inevitably requires numerous electric, electronic and mechanical quantities. In an attempt to highlight respective quantities that share similar characteristics in each of these fields, we try to term these identically in different fonts. 
Scalar- and vector-valued electric quantities are written in meager italic and blackboard fonts $a, \el a$ respectively, bold sans-serif fonts $\pt a$ is used for (vector- and tensor-valued) electronic quantities, and bold italic font $\gz a$ is selected for vector- and tensor-valued mechanical quantities. In order to facilitate the handling of this work, Table \ref{kinematic_expressions} summarizes the terms and expressions of the respective fields.

\begin{table}[!h]
\centering
 \begin{tabular}{C{0.3\textwidth} | C{0.4\textwidth}}
   \hline
 \multicolumn{2}{c}{Space-time quantities}\\
  \hline		
   material identity tensor   &{$\boldsymbol {I}$} \\
 spatial identity tensor & $\boldsymbol i$  \\
 material position vector  & $\boldsymbol X$\\
 spatial position vector  & $\boldsymbol x$\\
 time  & $t$\\
   \hline	
    \multicolumn{2}{c}{Electric quantities}\\
    \hline
       electric solution field   &{$y=y(\gz X, t)$} \\
 electric field & $\el E:=-\Grad y(\gz X, t)$  \\
 material rate of the electric solution field  & $v:=\D_t y(\gz X, t)$\\
    \hline	
    \multicolumn{2}{c}{Electronic quantities}\\
    \hline
       electronic solution field   &$\pt y^{\text{t}}, \pt y^{\text{c}}$ \\
 electronic order parameter field&  $\pt y:=\{\pt y^{\text t},\pt y^{\text c}\}$ \\
 space gradient of the electronic order parameter field & $\pt F:=\Grad\pt y(\gz X, t)$ \\
 material rate of the electronic order parameter field  & $\pt v:=\D_t\pt y(\gz X, t)$\\
     \hline	
    \multicolumn{2}{c}{Mechanical quantities}\\
    \hline
       mechanical solution field   &$\gz y$ \\
  deformation gradient & $\gz F:=\Grad\gz y(\gz X, t)$ \\
  Jacobian determinant & $ J=\text{det}(\gz F)$ \\
  cofactor & $ \gz K=\gz K(\gz F):=\cof\gz F=J \gz F^{-T}$ \\
  inverse deformation gradient & $\gz f:=\gz F^{-1}$ \\
  inverse Jacobian determinant & $ j=J^{-1}$ \\
  inverse cofactor & $ \gz k=\gz K^{-1}$ \\
 velocity  & $\gz v:=\D_t\gz y(\gz X, t)$\\
 \hline
\end{tabular} \caption{Summary of the necessary expressions for the description of a photo-mechanical modelling framework} \label{kinematic_expressions}
\end{table}


\section{Solution Fields}

Modelling light-matter interaction in photo-active polymers based on molecular photo-switches consists of coupled electric, electronic and mechanical sub-problems, each expressed in terms of a corresponding solution field.
Subsequently, we shall first briefly introduce the electric, electronic and mechanical solution fields together with their pertinent space-time gradients.

\subsection{Electric Problem}

The scalar-valued \textit{electric} solution field $y$, parameterized in terms of the material space coordinate $\gz X$ and time $t$, represents the electric potential
\be
y=y(\gz X, t)\with \el E:=-\Grad y(\gz X, t)\quad\mbox{and}\quad v:=\D_t y(\gz X, t).
\ee
Its (negative) material space gradient renders the nominal (Piola-type) electric field $\el E$, its material time gradient, introduced here merely for completeness, denotes the material rate $v$ of the electric potential.

\subsection{Electronic Problem}

The vector-valued \textit{electronic} solution fields $\pt y^{\rm t}$ and $\pt y^{\rm c}$, here collectively assembled in the double-vector-valued electronic order parameter field $\pt y:=\{\pt y^{\rm t},\pt y^{\rm c}\}$, represent the effective density of, e.g., vector-valued trans (low energy, rod shape) and cis (high energy, kinked shape) states of polymer-embedded photo-active azobenzene molecular switches
\be
\pt y=\pt y(\gz X, t)\with \pt F:=\Grad\pt y(\gz X, t)\und \pt v:=\D_t\pt y(\gz X, t).
\ee
The material space-time gradients $\pt F$ and $\pt v$ of the electronic order parameter field $\pt y$ capture its spatial and temporal changes.\\

\begin{figure}[ht!]
\centering
{\includegraphics[width=0.9\textwidth]{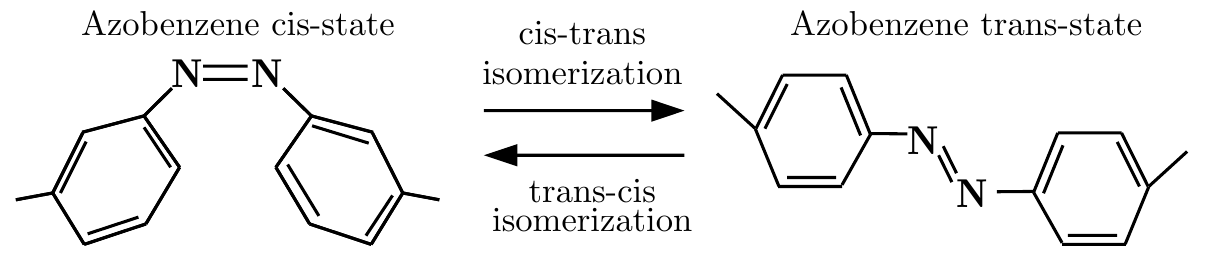}}
	\caption{Light induced transformation of an azobenzene molecule.}\label{Fig:azobenzene}
\end{figure}

In terms of generalized continua, the electronic order parameter(s) contained in $\pt y$ are attached to the material macro position vector $\gz X$. They may be thought of as effective micro position vectors (electronic coordinates) obtained by homogenizing micro position vectors connecting to photo-active charged particles of, e.g., azobenzene molecules within an RVE, see Figure \ref{Fig:azobenzene}.

\subsection{Mechanical Problem}

The vector-valued \textit{mechanical} solution field $\gz y$ represents the nonlinear deformation map of geometrically exact continuum kinematics. It maps material position vectors $\gz X$ (material coordinates) of physical points in the material (undeformed/reference) configuration into their spatial counterpart $\gz x$ in the spatial (deformed/current) configuration, i.e.\
\be
\gz x=\gz y(\gz X, t)\with \gz F:=\Grad\gz y(\gz X, t)\und \gz v:=\D_t\gz y(\gz X, t).
\ee
The corresponding material space-time gradients $\gz F$ and $\gz v$ render the deformation gradient (or rather the tangent map) and the velocity.\\

Regarding the deformation gradient $\gz F$ as the tangent map $\d\gz x=\gz F\cdot\d\gz X$ it proves convenient to introduce its cofactor $\gz K$ as the area map $\d\gz a=\gz K\cdot\d\gz A$ and its determinant $J$ as the volume map $\d v=J\d V$ via
\be
\gz K=\gz K(\gz F):=\cof\gz F=J \gz F^{-T}\und J=J(\gz F):=\det\gz F.
\ee

Moreover, it is useful to occasionally abbreviate the inverses of 
$\gz F$, $\gz K$, and $J$ as
\be
\gz f:=\gz F^{-1}\und \gz k=:\gz K^{-1}\und j:=J^{-1}.
\ee

In the sequel, the following derivatives of $\gz f, \gz K$ and $J$ are needed
\be
\frac{\partial\gz f}{\partial\gz F}=-\gz f\,\boxtimes\,\gz f^t\und
\frac{\partial\gz K}{\partial\gz F}=\gz f^t\otimes\gz K-\gz K\,\boxdot\,\gz f\with\frac{\partial J}{\partial\gz F}=\gz K,
\ee
whereby the non-standard dyadic products $\boxtimes$ and $\boxdot$ expand in Cartesian coordinate representation as $[\gz A\boxtimes\gz B]_{ijkl}:=A_{ik}B_{jl}$ and $[\gz A\boxdot\gz B]_{ijkl}:=A_{il}B_{jk}$.

\subsection{Re-Parameterizations}

Composition with the inverse deformation map $\gz X=\gz y^{-1}(\gz x, t)$ results in the re-parameterized electric solution field $\widetilde y$, i.e.\ the re-parameterized electric potential, and its corresponding (negative) spatial (space) gradient $\widetilde{\el e}$, i.e.\ the true (Cauchy-type) electric field
\be
\widetilde y(\gz x, t):=y(\gz X, t)\circ\gz y^{-1}\und\widetilde{\el e}:=-\grad\widetilde y(\gz x, t).
\ee

Likewise, composition with the inverse deformation map renders the re-parameterized electronic solution fields $\widetilde{\pt y}^{\rm t}$ and $\widetilde{\pt y}^{\rm c}$, i.e.\ the re-parameterized electronic order parameter field $\widetilde{\pt y}=\{\widetilde{\pt y}^{\rm t}, \widetilde{\pt y}^{\rm c}\}$, and its corresponding spatial (space) gradient $\widetilde{\pt f}$
\be
\widetilde{\pt y}(\gz x, t):=\pt y(\gz X, t)\circ\gz y^{-1}=\pt y(\gz y^{-1}(\gz x,t), t)\und\widetilde{\pt f}:=\grad\widetilde{\pt y}(\gz x, t).
\ee

Finally, composition of $\widetilde{\el e}$ and $\widetilde{\pt f}$ with the deformation map $\gz x=\gz y(\gz X, t)$ results in the re-parameterized spatial (space) gradients $\el e$ and $\pt f$, i.e.\\
\be
\el e=\el e(\gz X, t):=\widetilde{\el e}(\gz x, t)\circ\gz y=\el E\cdot\gz f\und
\pt f=\pt f(\gz X, t):=\widetilde{\pt f}(\gz x, t)\circ\gz y=\pt F\cdot\gz f.
\ee

Note the push-forward relation between $\el E$ and $\el e$ as well as between $\pt F$ and $\pt f$ in terms of the inverse deformation gradient $\gz f:=\gz F^{-1}$.\\

For the ease of notation we shall from here on use the sloppy notation $\el e$ and $\pt f$ also for $\widetilde{\el e}$ and $\widetilde{\pt f}$, thereby ignoring the parameterizations in either spatial or material coordinates $\gz x$ or $\gz X$, respectively, if there is no danger of confusion.

\section{Energy Densities}

Any of the variational settings as discussed below build on properly defined expressions for various kind of energy densities. We shall thus first discuss these separately for the electric, the electronic and the mechanical sub-problem. Thereby, we distinguish energy densities per unit volume in either the material or the spatial configuration by corresponding sub-scripts, i.e.\ the material $(\bullet)_{\rm m}$ versus the spatial $(\bullet)_{\rm s}$ density. These densities are related by the Jacobian $J$ of the deformation gradient as $(\bullet)_{\rm m}=J (\bullet)_{\rm s}$.

\subsection{Electric Problem}

The electric field penetrates free space and matter likewise, whereby electro-static energy is stored. We shall here denote the electro-statically stored energy as \textit{electric internal potential energy} with material density $e_{\rm m}=J e_{\rm s}$. Expressed in terms of the nominal electric field $\el E$ and the deformation gradient $\gz F$ (that in free space is a suited artificial extension of its counterpart in matter, see \cite{Bustamante2}), the electric internal potential energy density reads as
\be
e_{\rm m}=e_{\rm m}(\el E, \gz F):=
-\frac{1}{2}J\varepsilon_0\el e(\el E, \gz F)\cdot\el e(\el E, \gz F)=
-\frac{1}{2} \varepsilon_0\el e(\el E, \gz F)\cdot\gz K\cdot\el E=
-\frac{1}{2} \varepsilon_0\el E\cdot\gz f\cdot\gz K\cdot\el E.
\ee

Here, $\varepsilon_0$ denotes the electric permittivity of free space, a natural constant. Note i) the common quadratic expression of $2 e_{\rm s}=-\varepsilon_0\vert\el e\vert^2$ when expressed in terms of the true electric field $\el e$, and ii) the negative sign of $e_{\rm s}$ (thus a Legendre transformation $e_{\rm s}-\el e\cdot\partial_{\el e}e_{\rm s}$ results in a corresponding (dual) energy density with positive sign when expressed in terms of the conjugate variable $-\partial_{\el e}e_{\rm e}$).\\

For convenience of later analyses, we pre-compute the derivatives of the electric internal potential energy density $e_{\rm m}$ with respect to the nominal electric field $\el E$ and the deformation gradient $\gz F$ as
\be
-\frac{\partial e_{\rm m}}{\partial\el E}=                \varepsilon_0\el e             \cdot\gz K\und
 \frac{\partial e_{\rm m}}{\partial\gz F}=[e_{\rm s}\gz i+\varepsilon_0\el e\otimes\el e]\cdot\gz K,
\ee
where $\gz i$ is introduced as the spatial identity tensor. Observe the term $\varepsilon_0\el e$ representing the free space electric flux density (electric displacement) as well as the so-called energy-momentum format of $e_{\rm s}\gz i+\varepsilon_0\el e\otimes\el e$ representing the free space Maxwell stress (both of Cauchy-type).\\

Furthermore, we introduce the electric \textit{external potential energy} densities $v_{\rm m}^{\rm elec}$ and $\widehat v_{\rm m}^{\rm elec}$ in the bulk of matter and at the boundary between matter and free space, respectively, as
\be
v_{\rm m}^{\rm elec}=v_{\rm m}^{\rm elec}(y):=q^{\rm f}_{\rm m}\,y\und
\widehat v_{\rm m}^{\rm elec}=\widehat v_{\rm m}^{\rm elec}(y):=\widehat q^{\rm f}_{\rm m}\,y.
\ee

Here, $q^{\rm f}_{\rm m}$ and $\widehat q^{\rm f}_{\rm m}$ are the externally prescribed electric free charge densities per unit volume and unit area, respectively, in the material configuration.

\subsection{Electronic Problem}

The electronic solution fields $\pt y^{\rm t}$ and $\pt y^{\rm c}$ are associated with effective charge densities interacting with the electric field, thereby storing energy. For the sake of terminological consistency, we shall denote the corresponding stored energy as \textit{electronic internal potential energy} with density $c_{\rm m}=J c_{\rm s}$, it reads as
\be
c_{\rm m}=c_{\rm m}(\pt y, \el E, \gz F):=
-J\omega_0\pt y\cdot\el e(\el E, \gz F)=
- \omega_0\pt y\cdot\gz K\cdot\el E.
\ee

Here, $\omega_0=\{\omega_0^{\rm t},\omega_0^{\rm c}\}$ denote the effective charge densities bound to the electronic order parameter(s), whereby we shall assume $\omega_0$ as given and constant in order to avoid the necessity to include internal variables within a variational setting (we shall do so in a separate contribution). Different model options of time-varying effective charge densities are pursued in \cite{bin2015unified,oates2015non}.\\

For convenience of later analyses, we pre-compute the derivatives of the electronic internal potential energy density $c_{\rm m}$ with respect to the nominal electric field $\el E$, the electronic order parameter(s) $\pt y$ and the deformation gradient $\gz F$ as
\be
-\frac{\partial c_{\rm m}}{\partial\el E}= \omega_0\pt y\cdot\gz K\und
-\frac{\partial c_{\rm m}}{\partial\pt y}=J\omega_0\el e          \und
 \frac{\partial c_{\rm m}}{\partial\gz F}=[c_{\rm s}\gz i+\omega_0\el e\otimes\pt y]\cdot\gz K.
\ee

Observe the term $\omega_0\pt y$ as a contribution to the polarization in matter, the term $\omega_0\el e$ as an electronic (Lorentz-type) dipole force density in matter as well as the energy-momentum format of $c_{\rm s}\gz i+\omega_0\el e\otimes\pt y$ as a contribution of polarization to the Maxwell stress in matter.\\

Furthermore, we introduce the \textit{electronic external potential energy} densities $v_{\rm m}^{\rm tron}$ and $\widehat v_{\rm m}^{\rm tron}$ in the bulk of matter and at the boundary between matter and free space, respectively, as
\be
v_{\rm m}^{\rm tron}=v_{\rm m}^{\rm tron}(\pt y):=-\pt b_{\rm m}\cdot\pt y\und
\widehat v_{\rm m}^{\rm tron}=\widehat v_{\rm m}^{\rm tron}(\pt y):=-\pt t_{\rm m}\cdot\pt y.
\ee

Here, $\pt b_{\rm m}$ and $\pt t_{\rm m}$,  introduced for the sake of completeness, are externally prescribed electronic force densities per unit volume and unit area, respectively, in the material configuration.\\

The \textit{electronic kinetic energy} density $k_{\rm m}^{\rm tron}=J k_{\rm s}^{\rm tron}$ captures inertia of the electronic modes in terms of the material velocity $\pt v$ of the electronic order parameter(s) and
the electronic inertia density $\varrho_{\rm m}=J \varrho_{\rm s}$, a phenomenological parameter, as
\be
k_{\rm m}^{\rm tron}=k_{\rm m}^{\rm tron}(\pt v):=\frac{1}{2}\varrho_{\rm m}\pt v\cdot\pt v.
\ee


Finally, we introduce the \textit{electronic dissipation potential} density $\fr p_{\rm m}=J\fr p_{\rm s}$ in order 
to account for energy losses, e.g.\ due to optical scattering and/or photochemical reactions
\be
\fr p_{\rm m}=\fr p_{\rm m}(\pt v):=\frac{1}{2}J\gamma_0\pt v\cdot\pt v.
\ee

Here, $\gamma_0$ denotes a phenomenological damping parameter related to the material velocity $\pt v$ of the electronic order parameter(s). Its inverse relates to a time constant characterizing the relaxation of the material from a higher to a lower energetic excitation state.

\subsection{Mechanical Problem}

We shall denote the energy that is mechanically stored in matter as \textit{mechanical internal potential energy} with material density $w_{\rm m}=J w_{\rm s}$. Expressed in terms of the electronic order parameter(s) $\pt y$, their material gradient $\pt F$ and the deformation gradient $\gz F$, i.e.\ in terms of the micro and the macro deformation, the mechanical internal potential energy density reads generically as
\be
w_{\rm m}=w_{\rm m}(\pt y, \pt F, \gz F):=w_{\rm m}^\circ(\pt y, \gz F)+w_{\rm m}^\bullet(\pt F).
\ee

Here, we distinguish the contribution $w_{\rm m}^\circ$ due to the electronic order parameter(s) and the deformation gradient and the contribution $w_{\rm m}^\bullet$ due to the material gradient of the electronic order parameter(s). The former captures mechanically stored energy of e.g.\ amorphous azobenzene-polymer blends (as azobenzene-polyimide polymer networks), whereas the latter describes e.g.\ acrylate-based azobenzene-polymer blends displaying liquid crystal domain formation (as azobenzene-LCNs).\\

Furthermore, we introduce the \textit{mechanical external potential energy} densities $v_{\rm m}^{\rm mech}$ and $\widehat v_{\rm m}^{\rm mech}$ in the bulk of matter and at the boundary between matter and free space, respectively, as
\be
v_{\rm m}^{\rm mech}=v_{\rm m}^{\rm mech}(\gz y):=-\gz b_{\rm m}\cdot\gz y\und
\widehat v_{\rm m}^{\rm mech}=\widehat v_{\rm m}^{\rm mech}(\gz y):=-\gz t_{\rm m}\cdot\gz y.
\ee

Here, $\gz b_{\rm m}$ and $\gz t_{\rm m}$ are externally prescribed mechanical force densities per unit volume and unit area in the material configuration and $\widehat v_{\rm m}^{\rm mech}$ is given in  energy per area.\\

Finally, the \textit{mechanical kinetic energy} density $k_{\rm m}^{\rm mech}=J k_{\rm s}^{\rm mech}$ captures inertia of the mechanical modes in terms of the material velocity $\gz v$ of the deformation map and the mechanical inertia density $\rho_{\rm m}=J \rho_{\rm s}$ as
\be
k_{\rm m}^{\rm mech}=k_{\rm m}^{\rm mech}(\gz v):=\frac{1}{2}\rho_{\rm m}\gz v\cdot\gz v.
\ee


This concludes the energetic characterization of the electric, electronic and mechanical sub-problems.

\section{Constitutive Relations}

The derivatives of the various electric, electronic and mechanical energy densities with respect to their arguments (state variables) define constitutive expressions for their energetically conjugate quantities (state functions). Collectively, the state variables and state functions constitute the state quantities that describe the state of a system. We shall here introduce these constitutive relations as definitions for the convenience of later considerations, thereby distinguishing between nominal (Piola-type) and true (Cauchy-type) quantities.

\subsection{Electric Problem}

The nominal (Piola-type) electric flux density (or rather electric displacement) in free space, the nominal polarization and the nominal electric flux density in matter follow as
\be
\el D^\varepsilon:=-\frac{\partial e_{\rm m}}{\partial\el E}=\varepsilon_0\el e\cdot\gz K\und
\el P            :=-\frac{\partial c_{\rm m}}{\partial\el E}=     \omega_0\pt y\cdot\gz K\und
\el D            :=\el D^\varepsilon+\el P.
\ee

Moreover, these nominal quantities relate to the 
true (Cauchy-type) electric flux density in free space, the true polarization and the true electric flux density in matter via a Piola transformation, i.e.\ a right-sided push-forward with the inverse cofactor $\gz k$, to render
\be
\el d^\varepsilon:=-\frac{\partial e_{\rm s}}{\partial\el e}=\varepsilon_0\el e\und
\el p            :=-\frac{\partial c_{\rm s}}{\partial\el e}=     \omega_0\pt y\und
\el d            :=\el d^\varepsilon+\el p=\el D\cdot\gz k.
\ee

Finally, the nominal electric external source density, i.e.\ the free bulk charge, and the nominal electric external flux density, i.e. the free surface charge, derive form the electric external potential energy densities as
\be
         q^{\rm f}_{\rm m}:=\frac{\partial         v_{\rm m}^{\rm elec}}{\partial y}\und
\widehat q^{\rm f}_{\rm m}:=\frac{\partial\widehat v_{\rm m}^{\rm elec}}{\partial y}.
\ee

The free bulk charge and the free surface charge are here considered as given, i.e.\ as externally prescribed data. 

\subsection{Electronic Problem}

The electronic nominal (Piola-type) stress in matter follows as the derivative
\be
\pt P:=\frac{\partial w_{\rm m}}{\partial\pt F}=\frac{\partial w_{\rm m}^\bullet}{\partial\pt F}.
\ee

Moreover, it relates to the electronic true (Cauchy-type) stress via a Piola transformation to render
\be
\pt s:=\frac{\partial w_{\rm m}}{\partial\pt F}\cdot\gz k=\frac{\partial w_{\rm m}^\bullet}{\partial\pt F}\cdot\gz k.
\ee

Likewise, we define nominal energetic, dissipative, and total electronic internal source densities as
\be
\pt s_{\rm m}^\sharp:=\frac{\partial    w_{\rm m}}{\partial\pt y}=\frac{\partial w_{\rm m}^\circ}{\partial\pt y}\und
\pt s_{\rm m}^\flat :=\frac{\partial\fr p_{\rm m}}{\partial\pt v}=J\gamma_0\pt v\und
\pt s_{\rm m}       :=\pt s_{\rm m}^\sharp+\pt s_{\rm m}^\flat.
\ee

Then, their true counterparts, i.e.\ the true energetic, dissipative, and total electronic internal source densities compute as
\be
\pt s_{\rm s}^\sharp:=\frac{\partial    w_{\rm s}}{\partial\pt y}=\frac{\partial w_{\rm s}^\circ}{\partial\pt y}\und
\pt s_{\rm s}^\flat :=\frac{\partial\fr p_{\rm s}}{\partial\pt v}=\gamma_0\pt v\und
\pt s_{\rm s}       :=\pt s_{\rm s}^\sharp+\pt s_{\rm s}^\flat=j\pt s_{\rm m}.
\ee

Next, we introduce the nominal \textit{exterior} electronic external source density and the nominal electronic external flux density
\be
\pt b_{\rm m}^\blacktriangle:=-\frac{\partial         v_{\rm m}^{\rm tron}}{\partial\pt y}\und
\pt t_{\rm m}               :=-\frac{\partial\widehat v_{\rm m}^{\rm tron}}{\partial\pt y}.
\ee

The nominal exterior electronic external source density $\pt b_{\rm m}^\blacktriangle$ is complemented by its \textit{interior} counterpart $\pt b_{\rm m}^\vartriangle$ that is due to the coupling of the electronic order parameter(s) with the electric field. Collectively, these render the \textit{total} electronic external source density $\pt b_{\rm m}$, i.e.\
\be
\pt b_{\rm m}^\vartriangle:=-\frac{\partial         c_{\rm m}}{\partial\pt y}\und
\pt b_{\rm m}             :=\pt b_{\rm m}^\blacktriangle+\pt b_{\rm m}^\vartriangle.
\ee

The nominal \textit{exterior} electronic external source density and the nominal electronic external flux density are here considered as given, i.e.\ as externally prescribed data.\\

Finally, the nominal electronic momentum 
\be
\pt p_{\rm m}:=\frac{\partial k_{\rm m}^{\rm tron}}{\partial\pt v}=\varrho_{\rm m}\pt v
\ee
is the conjugate quantity to the material velocity $\pt v$ of the electronic order parameter(s).

\subsection{Mechanical Problem}

The electric and electronic (Maxwell stress) contributions to the nominal (or rather Piola) stress derive from the electric and electronic internal potential energy densities
\be
\gz P^{\rm elec}:=\frac{\partial e_{\rm m}}{\partial\gz F}=[e_{\rm s}\gz i+\el e\otimes\el d^\varepsilon]\cdot\gz K\und
\gz P^{\rm tron}:=\frac{\partial c_{\rm m}}{\partial\gz F}=[c_{\rm s}\gz i+\el e\otimes\el p            ]\cdot\gz K.
\ee

Likewise, the mechanical contribution to the nominal (or rather Piola) stress derives from the mechanical internal potential energy density
\be
\gz P^{\rm mech}:=\frac{\partial w_{\rm m}}{\partial\gz F}.
\ee

Collectively, these result in the total nominal (or rather Piola) stress as
\be
\gz P=\gz P^{\rm elec}+\gz P^{\rm tron}+\gz P^{\rm mech}=
\big[[e_{\rm s}+c_{\rm s}]\gz i+\el e\otimes\el d]\cdot\gz K+\frac{\partial w_{\rm m}}{\partial\gz F}.
\ee

Piola transformation then renders the electric and electronic contributions to the true (or rather Cauchy) stress
\be
\gz s^{\rm elec}:=\frac{\partial e_{\rm m}}{\partial\gz F}\cdot\gz k=[e_{\rm s}\gz i+\el e\otimes\el d^\varepsilon]\und
\gz s^{\rm tron}:=\frac{\partial c_{\rm m}}{\partial\gz F}\cdot\gz k=[c_{\rm s}\gz i+\el e\otimes\el p            ].
\ee

together with the mechanical contribution to the true (or rather Cauchy) stress
\be
\gz s^{\rm mech}:=\frac{\partial w_{\rm m}}{\partial\gz F}\cdot\gz k.
\ee

Again, collectively, these result in the total true (or rather Cauchy) stress
\be
\gz s=\gz s^{\rm elec}+\gz s^{\rm tron}+\gz s^{\rm mech}=
\big[[e_{\rm s}+c_{\rm s}]\gz i+\el e\otimes\el d]+\frac{\partial w_{\rm m}}{\partial\gz F}\cdot\gz k=\gz P\cdot\gz k.
\ee

Next, the nominal mechanical external source density, i.e.\ the volume-distributed body force, and the nominal mechanical external flux density, i.e.\ the area-distributed surface traction, derive as
\be
\gz b_{\rm m}:=-\frac{\partial         v_{\rm m}^{\rm mech}}{\partial\gz y}\und
\gz t_{\rm m}:=-\frac{\partial\widehat v_{\rm m}^{\rm mech}}{\partial\gz y}.
\ee

Finally, the nominal mechanical momentum is the conjugate quantity to the material velocity $\gz v$ of the deformation map
\be
\gz p_{\rm m}:=\frac{\partial k_{\rm m}^{\rm mech}}{\partial\gz v}=   \rho_{\rm m}\gz v.
\ee

This concludes the constitutive characterization of the electric, electronic and mechanical sub-problems.

\section{Dirichlet Principle}

For quasi-static situations without any inertia effects, Dirichlet's principle of stationary potential energy renders the pertinent equilibrium equations in the bulk and at the boundary, here for the electric, the electronic and the mechanical sub-problems. Traditionally, Dirichlet's principle is restricted to conservative, i.e.\ energetic cases void of dissipation. However, when expressed as an incremental variational problem in terms of the incremental work, also non-conservative, i.e.\ dissipative cases can be considered when properly incorporating a dissipation potential. In the sequel, we will demonstrate the variational setting of energetic and dissipative cases when modelling light-matter interaction in photo-active polymers.

\subsection{Energetic Case}

For the energetic case we first expand on the total potential energy densities before examining the pertinent variational setting.

\subsubsection{Potential Energy}

For the sake of convenience, we introduce the 
\textit{total potential energy} density $u_{\rm m}=J u_{\rm s}$ in the bulk of matter as the summation of the corresponding internal electric, electronic and  mechanical potential energies as well as the \textit{total external potential energy} density
\be
u_{\rm m}=u_{\rm m}(y, \pt y, \gz y, \el E, \pt F, \gz F):=
e_{\rm m}(\el E,\gz F)+c_{\rm m}(\pt y,\el E,\gz F)+w_{\rm m}(\pt y, \pt F, \gz F)+v_{\rm m}(y, \pt y, \gz y).
\ee

Thereby, the total external potential energy density $v_{\rm m}=J v_{\rm s}$ in the bulk of matter is the summation of the corresponding external electric, electronic and  mechanical potential energy densities
\be
v_{\rm m}=v_{\rm m}(y, \pt y, \gz y):=v_{\rm m}^{\rm elec}(y)+v_{\rm m}^{\rm tron}(\pt y)+v_{\rm m}^{\rm mech}(\gz y).
\ee


Moreover, we abbreviate the \textit{total external potential energy} density $\widehat v_{\rm m}=\widehat J\, \widehat v_{\rm s}$ at the boundary between matter and free space as the summation of the corresponding external electric, electronic and  mechanical potential energy densities
\be
\widehat v_{\rm m}=\widehat v_{\rm m}(y, \pt y, \gz y):=\widehat v_{\rm m}^{\rm elec}(y)+\widehat v_{\rm m}^{\rm tron}(\pt y)+\widehat v_{\rm m}^{\rm mech}(\gz y).
\ee

The total potential energy density $u_{\rm m}$ in the bulk of matter as well as the total external potential energy density $\widehat v_{\rm m}$ at the boundary between matter and free space together with the electric internal potential energy density $e_{\rm m}$ in the bulk of free space contribute to the potential energy functional as discussed in the sequel.

\subsubsection{Variational Setting}

To begin with, we define the \textit{potential energy functional} for Dirichlet's principle as 
\bea
U=U(y, \pt y, \gz y)&:=&\displaystyle
\int_{\cl B_{\rm m}}u_{\rm m}(y, \pt y, \gz y, \el E, \pt F, \gz F)\d V\nonumber\\&+&\displaystyle
\int_{\partial\cl B_{\rm m}}\widehat v_{\rm m}(y, \pt y, \gz y)\d A\\\nonumber&+&\displaystyle
\int_{\cl S_{\rm m}}e_{\rm m}(\el E, \gz F)\d V.
\eea

Then, Dirichlet's principle requires stationarity of the potential energy functional upon admissible, i.e.\ (space) boundary conditions satisfying material variation $\D_\delta$ (i.e.\ variation at fixed material position $\gz X$) of the solution fields $y, \pt y, \gz y$ as
\be
U(y, \pt y, \gz y)\to\mbox{stationary point}.
\ee

Concretely, the stationarity condition for the potential energy functional expands as
\bea{}
\D_\delta U&=&\displaystyle\int_{\cl B_{\rm m}}\D_\delta u_{\rm m}(y, \pt y, \gz y, \el E, \pt F, \gz F)\d V\nonumber\\&+&\displaystyle
\int_{\partial\cl B_{\rm m}}\D_\delta\widehat v_{\rm m}(y, \pt y, \gz y)\d A\\\nonumber&+&\displaystyle
\int_{\cl S_{\rm m}}\D_\delta e_{\rm m}(\el E, \gz F)\d V\doteq 0\quad\forall\D_\delta y, \D_\delta\pt y, \D_\delta\gz y.
\eea

Requiring stationarity of the potential energy functional for arbitrary admissible $\D_\delta y, \D_\delta\pt y, \D_\delta\gz y$ and using the constitutive relations as introduced in the above, results eventually in the Euler-Lagrange or rather \textit{equilibrium equations}\footnote{
Based on the dependency of the potential energy functional on the solution fields and their material space gradients, variational calculus results in the following Euler-Lagrange equations:\\

$\bullet$ Euler-Lagrange equations in the bulk of matter
\ben
\Div\frac{\partial u_{\rm m}}{\partial\Grad    y}=\frac{\partial u_{\rm m}}{\partial    y},\qquad
\Div\frac{\partial u_{\rm m}}{\partial\Grad\pt y}=\frac{\partial u_{\rm m}}{\partial\pt y},\qquad
\Div\frac{\partial u_{\rm m}}{\partial\Grad\gz y}=\frac{\partial u_{\rm m}}{\partial\gz y}.
\een

$\bullet$ Euler-Lagrange equations in the bulk of free space
\ben
\Div\frac{\partial e_{\rm m}}{\partial\Grad    y}=    0,\qquad
\Div\frac{\partial e_{\rm m}}{\partial\Grad\gz y}=\tz 0.
\een

$\bullet$ Euler-Lagrange equations at the boundary between matter and free space
\ben
\left[\frac{\partial u_{\rm m}}{\partial\Grad    y}-
      \frac{\partial e_{\rm m}}{\partial\Grad    y}\right]\cdot\gz N=-\frac{\partial\widehat v_{\rm m}}{\partial    y},\qquad
      \frac{\partial u_{\rm m}}{\partial\Grad\pt y}       \cdot\gz N=-\frac{\partial\widehat v_{\rm m}}{\partial\pt y},\qquad
\left[\frac{\partial u_{\rm m}}{\partial\Grad\gz y}-
      \frac{\partial e_{\rm m}}{\partial\Grad\gz y}\right]\cdot\gz N=-\frac{\partial\widehat v_{\rm m}}{\partial\gz y}
\een
Identifying the individual terms with the constitutive relations as in the above renders the result.
}
\begin{framed}
\begin{subequations}
\bea
\Div\el D-q^{\rm f}_{\rm m}      \,=   0\;\:\:\quad\mbox{in}\:\!\quad\cl B_{\rm m}&\mbox{and}&\:\:\:\jmp{\el D}  \cdot\gz N= \widehat q^{\rm f}_{\rm m}\!\!\,\quad\mbox{at}\quad\partial\cl B_{\rm m}\\
\Div\el D\hspace{9.7mm}\,=   0\;\:\:\quad\mbox{in}\:\!\quad\cl S_{\rm m}&          &\\[2mm]
\Div\pt P\!\;+\pt b_{\rm m}    \!\:=\pt s_{\rm m}^\sharp\quad\mbox{in}\quad\cl B_{\rm m}&\mbox{and}&\hspace{5mm}\,\pt P\:\cdot\gz N=\pt t_{\rm m}\quad\;\!\!\!\mbox{at}\quad\partial\cl B_{\rm m}\\[2mm]
\Div\gz P+\gz b_{\rm m}      \,=\tz   0\;\;\quad\mbox{in}\quad\cl B_{\rm m}&\mbox{and}&-\jmp{\gz P}  \cdot\gz N=\gz t_{\rm m}\!\quad\mbox{at}\quad\partial\cl B_{\rm m}\\
\Div\gz P\hspace{9.9mm}\,=\tz   0\;\;\quad\mbox{in}\quad\cl S_{\rm m}&          &
\eea
\end{subequations}
\end{framed}

Observe that the Neumann-type boundary condition for the mechanical sub-problem appears as jump condition for the total Piola stress $\gz P$ at the boundary between matter and free space, thus involving the Maxwell stress $\gz P^{\rm elec}$ as present in the free space and exerted on the continuum body as the corresponding Maxwell traction \cite{Mehnert1,Vu}. Consequently, for polymers with low relative permittivity $\varepsilon_{\rm r}$ in the order of some $10^0-10^1$, the free space sub-problem is indeed non-negligible.\\

Furthermore, note that the equilibrium equations follow in terms of flux and source densities per unit area and volume, respectively, in the material configuration.
For completeness, Piola transformation then renders the entirely equivalent expressions in terms of flux and source densities per unit area and volume, respectively, in the spatial configuration
\begin{subequations}
\bea
\div\el d-q^{\rm f}_{\rm s}      \,=   0\;\:\:\quad\mbox{in}\:\!\quad\cl B_{\rm s}&\mbox{and}&\:\:\:\jmp{\el d}  \cdot\gz n= \widehat q^{\rm f}_{\rm s}\!\!\,\quad\mbox{at}\quad\partial\cl B_{\rm s}\\
\div\el d\hspace{8.8mm}\,=   0\;\:\:\quad\mbox{in}\:\!\quad\cl S_{\rm s}&          &\\[2mm]
\div\pt s\!\;+\pt b_{\rm s}    \:=\pt s_{\rm s}^\sharp\;\quad\mbox{in}\quad\cl B_{\rm s}&\mbox{and}&\hspace{5mm}\,\pt s\:\cdot\gz n=\pt t_{\rm s}\quad\;\!\!\!\mbox{at}\quad\partial\cl B_{\rm s}\\[2mm]
\div\gz s\,+\gz b_{\rm s}     \:=\tz   0\;\;\quad\mbox{in}\quad\cl B_{\rm s}&\mbox{and}&-\jmp{\gz s}  \cdot\gz n=\gz t_{\rm s}\!\!\:\quad\mbox{at}\quad\partial\cl B_{\rm s}\\
\div\gz s\hspace{9.7mm}\,=\tz   0\;\;\quad\mbox{in}\quad\cl S_{\rm s}&          &
\eea
\end{subequations}

Expanding in particular the term $\div\gz s$ in the equilibrium equation related to the deformation map in the bulk of matter renders\footnote{
A step-by-step derivation using $\grad\el e\cdot\el d-\el d\cdot\grad\el e=\rot\,\el e\times\el d=\el o$ and thus likewise $\grad\el e\cdot\el p-\el p\cdot\grad\el e=\el o$ is
\bean
\div\gz s&=&
\div\gz s^{\rm mech}-\el d\cdot\grad\el e-\el e\cdot\grad\el p+\grad\el e\cdot\el d+q_{\rm s}^{\rm f}\,\el e,\\\nonumber&=&
\div\gz s^{\rm mech}+\rot\,\el e\times\el d-\el e\cdot\grad\el p+q_{\rm s}^{\rm f}\,\el e,\\\nonumber&=&
\div\gz s^{\rm mech}+\grad\el e\cdot\el p-[\el p\cdot\grad\el e+\el e\cdot\grad\el p]+q_{\rm s}^{\rm f}\,\el e,\\\nonumber&=&
\div\gz s^{\rm mech}-\grad(\el p\cdot\el e)+\grad\el e\cdot\el p+q_{\rm s}^{\rm f}\,\el e,\\\nonumber&=&
\div(c_{\rm s}\gz i+\gz s^{\rm mech})+\grad\el e\cdot\el p+q_{\rm s}^{\rm f}\,\el e.
\eean
}
\be
\div\gz s=\div(c_{\rm s}\gz i+\gz s^{\rm mech})+\grad\el e\cdot\el p+q_{\rm s}^{\rm f}\,\el e
\ee

thereby clearly identifying the classical Lorentz-type volume forces $\grad\el e\cdot\el p+q_{\rm s}^{\rm f}\el e$ \cite{pelteret2019magneto,dorfmann1} due to the bound and free charge densities together with an additional pressure-like term $c_{\rm s}\gz i$ in the Cauchy stress that is due to the additional solution field or rather electronic order parameter(s) $\pt y$.\\

The equilibrium equations, expressed in terms of flux and source densities per unit area and volume, respectively, in either the material or the material configuration complete the variational setting of the energetic case for quasi-static situations.\\ 

Obviously, which of the equivalent alternative versions is used for solving coupled boundary value problems of light-matter interaction in photo-active polymers is largely a matter of taste.

\subsection{Dissipative Case}

For the dissipative case we consider incremental work densities as basis ingredients for the pertinent variational setting that allows inclusion of a dissipation potential.

\subsubsection{Incremental Work}

For an extension towards the dissipative case, we first introduce and abbreviate the increments of the electric, electronic and mechanical solution fields as
\be
w:=\d y,\qquad\pt w:=\d\pt y,\qquad\gz w:=\d\gz y.
\ee

Then, the \textit{incremental work density} $\fr u_{\rm m}=J \fr u_{\rm s}$ in the bulk of matter follows as the increment $\d u_{\rm m}$ of the total potential energy density  evaluated at \textit{fixed configuration space} $y, \pt y, \gz y$ as
\be
\fr u_{\rm m}(w, \pt w, \gz w, \Grad w, \Grad\pt w, \Grad\gz w):=\big[\d u_{\rm m}(y, \pt y, \gz y, \el E, \pt F, \gz F)\,\big]_{\rm fixed\: configuration\: space}.
\ee

As a result, and incorporating the constitutive relations as introduced in the above, the explicit representation of the incremental work density in the bulk of matter follows as
\bea{}
&\fr u_{\rm m}(w, \pt w, \gz w, \Grad w, \Grad\pt w, \Grad\gz w)=&\\\nonumber
&[\el D\cdot\Grad w+q^{\rm f}_{\rm m}\, w]+[\pt P:\Grad\pt w-[\pt b_{\rm m}-\pt s_{\rm m}^\sharp]\cdot\pt w]+[\gz P:\Grad\gz w-\gz b_{\rm m}\cdot\gz w].&
\eea

Correspondingly, the external incremental work density $\widehat{\fr v}_{\rm m}=\widehat J\, \widehat{\fr v}_{\rm s}$ at the boundary between matter and free space computes at fixed configuration space $y, \pt y, \gz y$ as
\be{}
\widehat{\fr v}_{\rm m}(w, \pt w, \gz w):=\big[\d\widehat v_{\rm m}(y, \pt y, \gz y)\,\big]_{\rm fixed\: configuration\: space}=
\widehat q^{\rm f}_{\rm m}\, w-\pt t_{\rm m}\cdot\pt w-\gz t_{\rm m}\cdot\gz w.
\ee

Finally, the electric incremental work density $\fr e_{\rm m}=J \fr e_{\rm s}$ in the bulk of free space expands at fixed configuration space $y,\gz y$ as
\be{}
\fr e_{\rm m}(\Grad w, \Grad\gz w):=\big[\d e_{\rm m}(\el E, \gz F)\,\big]_{\rm fixed\: configuration\: space}=
\el D\cdot\Grad w+\gz P:\Grad\gz w.
\ee

The total incremental work density $\fr u_{\rm m}$ in the bulk of matter as well as the external incremental work density $\widehat{\fr v}_{\rm m}$ at the boundary between matter and free space together with the electric incremental work density $\fr e_{\rm m}$ in the bulk of free space contribute to the incremental work functional as discussed in the sequel.

\subsubsection{Variational Setting}

The \textit{incremental work functional} allows inclusion of a dissipation potential $\fr p_{\rm m}$ and reads
\bea
\fr U=\fr U(w, \pt w, \gz w)&:=&
\int_{\cl B_{\rm m}}[\fr u_{\rm m}(w, \pt w, \gz w, \Grad w, \Grad\pt w, \Grad\gz w)+\fr p_{\rm m}(\pt w/\d t)\d t]\d V\nonumber\\&+&\displaystyle
\int_{\partial\cl B_{\rm m}}\widehat{\fr v}_{\rm m}(w, \pt w, \gz w)\d A\\\nonumber&+&\displaystyle
\int_{\cl S_{\rm m}}\fr e_{\rm m}(\Grad w, \Grad\gz w)\d V.
\eea

Note that in order to obtain an incremental quantity, the dissipation potential $\fr p_{\rm m}$ (which is a power-like quantity of dimension incremental work density per time) is multiplied by $\d t$, whereby based on the increment $\pt w:=\d\pt y$ its argument expresses as $\pt v:=\d\pt y/\d t\equiv\pt w/\d t$. Recall that here, i.e.\ in quasi-static situations, time is merely a parameter that orders the sequence of external loading.\\

Then, the incremental Dirichlet principle requires stationarity of the incremental work functional upon admissible, i.e.\ (space) boundary conditions satisfying material variation $\D_\delta$ of the incremental solution fields $w, \pt w, \gz w$ as
\be
\fr U(w, \pt w, \gz w)\to\mbox{stationary point}.
\ee

Concretely, the stationarity condition for the incremental work functional expands as
\bea{}
\D_\delta\fr U&=&\displaystyle
\int_{\cl B_{\rm m}}[\D_\delta\fr u_{\rm m}(w, \pt w, \gz w, \Grad w, \Grad\pt w, \Grad\gz w)+\D_\delta\fr p_{\rm m}(\pt w/\d t)\d t]\d V\nonumber\\&+&\displaystyle
\int_{\partial\cl B_{\rm m}}\D_\delta\widehat{\fr v}_{\rm m}(w, \pt w, \gz w)\d A\\
\nonumber &+&\displaystyle
\int_{\cl S_{\rm m}}\D_\delta\fr e_{\rm m}(\Grad w, \Grad\gz w)\d V\doteq 0\quad\forall\D_\delta w, \D_\delta\pt w, \D_\delta\gz w.
\eea

Requiring stationarity of the incremental work functional for arbitrary admissible $\D_\delta w, \D_\delta\pt w, \D_\delta\gz w$ and using the explicit expressions for the incremental work densities as introduced in the above results eventually in the Euler-Lagrange or rather equilibrium equations\footnote{
Based on the dependency of the incremental work functional on the incremental solution fields and their material space gradients, variational calculus results in the following Euler-Lagrange equations:\\

$\bullet$ Euler-Lagrange equations in the bulk of matter
\ben
\Div\el D=q^{\rm f}_{\rm m},\qquad
\Div\pt P=\pt s_{\rm m}-\pt b_{\rm m},\qquad
\Div\gz P=-\gz b_{\rm m}.
\een

$\bullet$ Euler-Lagrange equations in the bulk of free space
\ben
\Div\el D=    0,\qquad
\Div\gz P=\tz 0.
\een

$\bullet$ Euler-Lagrange equations at the boundary between matter and free space
\ben
\jmp{\el D}\cdot\gz N=\widehat q_{\rm m}^{\rm f},\qquad
     \pt P \cdot\gz N=     \pt t_{\rm m},\qquad
\jmp{\gz P}\cdot\gz N=-    \gz t_{\rm m}.
\een
}
\begin{framed}
\begin{subequations}
\bea
\Div\el D-q^{\rm f}_{\rm m}      \,=   0\;\:\:\quad\mbox{in}\:\!\quad\cl B_{\rm m}&\mbox{and}&\:\:\:\jmp{\el D}  \cdot\gz N= \widehat q^{\rm f}_{\rm m}\!\!\,\quad\mbox{at}\quad\partial\cl B_{\rm m}\\
\Div\el D\hspace{9.7mm}\,=   0\;\:\:\quad\mbox{in}\:\!\quad\cl S_{\rm m}&          &\\[2mm]
\Div\pt P\!\;+\pt b_{\rm m}    \!\:=\pt s_{\rm m}\quad\mbox{in}\quad\cl B_{\rm m}&\mbox{and}&\hspace{5mm}\,\pt P\:\cdot\gz N=\pt t_{\rm m}\;\!\!\!\quad\mbox{at}\quad\partial\cl B_{\rm m}\\[2mm]
\Div\gz P+\gz b_{\rm m}      \,=\tz   0\;\;\quad\mbox{in}\quad\cl B_{\rm m}&\mbox{and}&-\jmp{\gz P}  \cdot\gz N=\gz t_{\rm m}\!\quad\mbox{at}\quad\partial\cl B_{\rm m}\\
\Div\gz P\hspace{9.9mm}\,=\tz   0\;\;\quad\mbox{in}\quad\cl S_{\rm m}&          &
\eea
\end{subequations}
\end{framed}

Expanding the equilibrium equation related to the electronic order parameter(s) (micro deformation) in the bulk of matter and neglecting external electronic sources, eventually allows term-by-term comparison with the formulation outlined in \cite{bin2015unified}

\be
\Div      \frac{\partial w^\bullet_{\rm m}}{\partial\pt F}                +J\omega_0\el e=
\frac{\partial w^\circ_{\rm m}}{\partial\pt y}+J\gamma_0\D_t\pt y\lto
\div\bigg(\frac{\partial w^\bullet_{\rm m}}{\partial\pt F}\cdot\gz k\bigg)+ \omega_0\el e=
\frac{\partial w^\circ_{\rm s}}{\partial\pt y}+ \gamma_0\D_t\pt y.
\ee
Thereby, the first terms left and right capture electronic forces associated with structured and amorphous regions, respectively, whereas the second terms on left and right describe the electronic (Lorentz-type) dipole force density in matter and the energy losses, e.g. due to optical scattering and/or photochemical reactions, respectively.\\

For completeness, Piola transformation of the equilibrium equations then renders the equivalent expressions in terms of flux and source densities per unit area and volume, respectively, in the spatial configuration
\begin{subequations}
\bea
\div\el d-q^{\rm f}_{\rm s}      \,=   0\;\:\:\quad\mbox{in}\:\!\quad\cl B_{\rm s}&\mbox{and}&\:\:\:\jmp{\el d}  \cdot\gz n= \widehat q^{\rm f}_{\rm s}\!\!\,\quad\mbox{at}\quad\partial\cl B_{\rm s}\\
\div\el d\hspace{8.8mm}\,=   0\;\:\:\quad\mbox{in}\:\!\quad\cl S_{\rm s}&          &\\[2mm]
\div\pt s\!\;+\pt b_{\rm s}    \:=\pt s_{\rm s}\;\quad\mbox{in}\quad\cl B_{\rm s}&\mbox{and}&\hspace{5mm}\,\pt s\:\cdot\gz n=\pt t_{\rm s}\quad\;\!\!\!\mbox{at}\quad\partial\cl B_{\rm s}\\[2mm]
\div\gz s\,+\gz b_{\rm s}     \:=\tz   0\;\;\quad\mbox{in}\quad\cl B_{\rm s}&\mbox{and}&-\jmp{\gz s}  \cdot\gz n=\gz t_{\rm s}\!\!\:\quad\mbox{at}\quad\partial\cl B_{\rm s}\\
\div\gz s\hspace{9.7mm}\,=\tz   0\;\;\quad\mbox{in}\quad\cl S_{\rm s}&          &
\eea
\end{subequations}

Note the total electronic internal source
density $\pt s_{\rm s}:=\pt s_{\rm s}^\sharp+\pt s_{\rm s}^\flat$, consisting of energetic \textit{and} dissipative contributions, that appears in the equilibrium equation related to the electronic order parameter(s).\\

This concludes derivation of the equilibrium equations for the electric, electronic and mechanical sub-problems, embracing energetic as well as dissipative cases, from Dirichlet's principle.

\section{Hamilton Principle}

For dynamic situations, Hamilton's principle of least action states that the dynamics of a system between two given points in time, captured by the evolution of the state space coordinates, renders the action integral, i.e.\ a functional over the state space, a stationary value upon material variations of the state space coordinates. In the sequel, we will demonstrate the variational setting and the ensuing balance equations for energetic and dissipative cases when modelling light-matter interaction in photo-active polymers.

\subsection{Energetic Case}

For the energetic case we first expand on the total Lagrangian energy density before examining the pertinent variational setting.

\subsubsection{Lagrangian Energy}

We introduce the \textit{total Lagrangian energy density} $l_{\rm m}=J l_{\rm s}$ in the bulk of matter as the difference between the \textit{total kinetic energy density} $k_{\rm m}$ and the total potential energy density $u_{\rm m}$, thus
\be
l_{\rm m}=l_{\rm m}(y, \pt y, \gz y, \el E, \pt F, \gz F, \pt v, \gz v):=k_{\rm m}(\pt v, \gz v)-u_{\rm m}(y, \pt y, \gz y, \el E, \pt F, \gz F).
\ee

Thereby, the total kinetic energy density $k_{\rm m}=J k_{\rm s}$ in the bulk of matter consists of electronic and mechanical contributions
\be
k_{\rm m}=k_{\rm m}(\pt v, \gz v):=k_{\rm m}^{\rm tron}(\pt v)+k_{\rm m}^{\rm mech}(\gz v).
\ee

The total Lagrangian energy density $l_{\rm m}$ in the bulk of matter as well as the (negative) total external potential energy density $\widehat v_{\rm m}$ at the boundary between matter and free space together with the (negative) electric internal potential energy density $e_{\rm m}$ in the bulk of free space contribute to the action functional as discussed in the sequel.

\subsubsection{Variational Setting}

To begin with, we define the \textit{action functional} for Hamilton's principle as
\bea
A=A(y, \pt y, \gz y)&:=&\int_{\cl T}\bigg[\displaystyle
\int_{\cl B_{\rm m}}l_{\rm m}(y, \pt y, \gz y, \el E, \pt F, \gz F, \pt v, \gz v)\d V\nonumber\\&&\displaystyle\quad\;-\!
\int_{\partial\cl B_{\rm m}}\widehat v_{\rm m}(y, \pt y, \gz y)\d A\\\nonumber&&\displaystyle\quad\;-\!
\int_{\cl S_{\rm m}}e_{\rm m}(\el E, \gz F)\d V\bigg]\d t.
\eea

Then, Hamilton's principle requires stationarity of the action functional upon admissible, i.e.\ space-time boundary conditions satisfying material variation $\D_\delta$ of the solution fields $y, \pt y, \gz y$ as
\be
A(y, \pt y, \gz y)\to\mbox{stationary point}.
\ee

Concretely, the stationarity condition for the action functional expands as
\bea{}
\D_\delta A&=&\int_{\cl T}\bigg[\displaystyle\int_{\cl B_{\rm m}}\D_\delta l_{\rm m}(y, \pt y, \gz y, \el E, \pt F, \gz F, \pt v, \gz v)\d V\nonumber\\&&\displaystyle\quad\;-\!
\int_{\partial\cl B_{\rm m}}\D_\delta\widehat v_{\rm m}(y, \pt y, \gz y)\d A\\\nonumber&&\displaystyle\quad\;-\!
\int_{\cl S_{\rm m}}\D_\delta e_{\rm m}(\el E, \gz F)\d V\bigg]\d t\doteq 0\quad\forall\D_\delta y, \D_\delta\pt y, \D_\delta\gz y.
\eea

Requiring stationarity of the action functional for arbitrary admissible $\D_\delta y, \D_\delta\pt y, \D_\delta\gz y$ and using the constitutive relations as introduced in the above, results eventually in the Euler-Lagrange or rather \textit{balance equations}\footnote{
Based on the dependency of the action functional on the solution fields and their material space-time gradients, variational calculus results in the following Euler-Lagrange equations:\\

$\bullet$ Euler-Lagrange equations in the bulk of matter
\ben
0                                           =
\frac{\partial l_{\rm m}}{\partial    y}-\Div\frac{\partial l_{\rm m}}{\partial\Grad    y},\qquad
\D_t\frac{\partial l_{\rm m}}{\partial\pt v}=
\frac{\partial l_{\rm m}}{\partial\pt y}-\Div\frac{\partial l_{\rm m}}{\partial\Grad\pt y},\qquad
\D_t\frac{\partial l_{\rm m}}{\partial\gz v}=
\frac{\partial l_{\rm m}}{\partial\gz y}-\Div\frac{\partial l_{\rm m}}{\partial\Grad\gz y}.
\een

$\bullet$ Euler-Lagrange equations in the bulk of free space
\ben
    0=-\Div\frac{\partial e_{\rm m}}{\partial\Grad    y},\qquad
\tz 0=-\Div\frac{\partial e_{\rm m}}{\partial\Grad\gz y}.
\een

$\bullet$ Euler-Lagrange equations at the boundary between matter and free space
\ben
-\left[\frac{\partial e_{\rm m}}{\partial\Grad    y}+
       \frac{\partial l_{\rm m}}{\partial\Grad    y}\right]\cdot\gz N=-\frac{\partial\widehat v_{\rm m}}{\partial    y},\qquad
-
       \frac{\partial l_{\rm m}}{\partial\Grad\pt y}       \cdot\gz N=-\frac{\partial\widehat v_{\rm m}}{\partial\pt y},\qquad
-\left[\frac{\partial e_{\rm m}}{\partial\Grad\gz y}+
       \frac{\partial l_{\rm m}}{\partial\Grad\gz y}\right]\cdot\gz N=-\frac{\partial\widehat v_{\rm m}}{\partial\gz y}
\een
}
\begin{framed}
\begin{subequations}
\bea
\Div\el D-q^{\rm f}_{\rm m}      \,=   0\;\:\:\;\;\,\qquad\quad\quad\mbox{in}\:\!\quad\cl B_{\rm m}&\mbox{and}&\:\:\:\jmp{\el D}  \cdot\gz N= \widehat q^{\rm f}_{\rm m}\!\!\,\quad\mbox{at}\quad\partial\cl B_{\rm m}\\
\Div\el D\hspace{9.7mm}\,=   0\;\:\:\;\;\,\qquad\quad\quad\mbox{in}\:\!\quad\cl S_{\rm m}&          &\\[2mm]
\Div\pt P\!\;+\pt b_{\rm m}    \!\:=\pt s_{\rm m}^\sharp+\D_t\pt p_{\rm m}\quad\mbox{in}\quad\cl B_{\rm m}&\mbox{and}&\hspace{5mm}\,\pt P\:\cdot\gz N=\pt t_{\rm m}\;\!\!\!\quad\mbox{at}\quad\partial\cl B_{\rm m}\\[2mm]
\Div\gz P+\gz b_{\rm m}      \,=\,\,\qquad\D_t\gz p_{\rm m}\quad\mbox{in}\quad\cl B_{\rm m}&\mbox{and}&-\jmp{\gz P}  \cdot\gz N=\gz t_{\rm m}\!\quad\mbox{at}\quad\partial\cl B_{\rm m}\\
\Div\gz P\hspace{9.9mm}\,=\tz   0\;\qquad\qquad\quad\mbox{in}\quad\cl S_{\rm m}&          &
\eea
\end{subequations}
\end{framed}

For completeness, Piola transformation then renders the equivalent expressions in terms of momentum, flux and source densities per unit area and volume, respectively, in the spatial configuration

\begin{subequations}
\bea
\div\el d-q^{\rm f}_{\rm s}      \,=   0\;\:\:\:\;\;\,\qquad\quad\quad\mbox{in}\:\!\quad\cl B_{\rm s}&\mbox{and}&\:\:\:\jmp{\el d}  \cdot\gz n= \widehat q^{\rm f}_{\rm s}\!\!\,\quad\mbox{at}\quad\partial\cl B_{\rm s}\\
\div\el d\hspace{8.8mm}\,=   0\;\;\;\;\;\;\qquad\quad\quad\mbox{in}\:\!\quad\cl S_{\rm s}&          &\\[2mm]
\div\pt s\!\;+\pt b_{\rm s}    \:=\pt s_{\rm s}^\sharp+j\D_t\pt p_{\rm m}\quad\mbox{in}\quad\cl B_{\rm s}&\mbox{and}&\hspace{5mm}\,\pt s\:\cdot\gz n=\pt t_{\rm s}\quad\;\!\!\!\mbox{at}\quad\partial\cl B_{\rm s}\\[2mm]
\div\gz s\,+\:\!\gz b_{\rm s}     \;=\qquad j\D_t\gz p_{\rm m}\quad\mbox{in}\quad\cl B_{\rm s}&\mbox{and}&-\jmp{\gz s}  \cdot\gz n=\gz t_{\rm s}\!\!\:\quad\mbox{at}\quad\partial\cl B_{\rm s}\\
\div\gz s\hspace{9.7mm}\,=\tz   0\;\,\qquad\qquad\quad\mbox{in}\quad\cl S_{\rm s}&          &
\eea
\end{subequations}

Note the additional inertia contributions for the electronic and the mechanical sub-problem as compared to the quasi-static situation.

\subsection{Dissipative Case}

For the dissipative case we consider the incremental action density as basis ingredient for the pertinent variational setting that allows inclusion of a dissipation potential.

\subsubsection{Incremental Action}

For an extension towards the dissipative case, we first introduce and abbreviate the velocities (material time derivatives) of the incremental electronic and mechanical solution fields as
\be
\D_t\pt w:=\D_t\d\pt y=\d\D_t\pt y\equiv\d\pt v,\qquad\D_t\gz w:=\D_t\d\gz y=\d\D_t\gz y\equiv\d\gz v.
\ee

The equivalence between the velocities of the increments and the increments of the velocities (of the electronic and mechanical solution fields) relies on the commutativity $\D_t\d(\bullet)\equiv\d\D_t(\bullet)$ of material time derivatives and increments.\\

Then, the \textit{incremental action density} $\fr l_{\rm m}=J \fr l_{\rm s}$ in the bulk of matter follows as the increment $\d l_{\rm m}$ of the total Lagrangian energy density  evaluated at \textit{fixed state space} $y, \pt y, \gz y, \pt v, \gz v$ as
\be
\fr l_{\rm m}(w, \pt w, \gz w, \Grad w, \Grad\pt w, \Grad\gz w, \D_t\pt w, \D_t\gz w):=\big[\d l_{\rm m}(y, \pt y, \gz y, \el E, \pt F, \gz F, \pt v, \gz v)\,\big]_{\rm fixed\: state\: space}.
\ee

As a result, and incorporating the constitutive relations as introduced in the above, the explicit representation of the incremental action density in the bulk of matter follows as
\bea{}
&\fr l_{\rm m}(w, \pt w, \gz w, \Grad w, \Grad\pt w, \Grad\gz w, \D_t\pt w, \D_t\gz w)=&\\\nonumber
&-[\el D\cdot\Grad w+q^{\rm f}_{\rm m}\, w]-[\pt P:\Grad\pt w-[\pt b_{\rm m}-\pt s_{\rm m}^\sharp]\cdot\pt w-\pt p_{\rm m}\cdot\D_t\pt w]-[\gz P:\Grad\gz w-\gz b_{\rm m}\cdot\gz w-\gz p_{\rm m}\cdot\D_t\gz w].&
\eea

The incremental action density $\fr l_{\rm m}$ in the bulk of matter as well as the (negative) external incremental work density $\widehat{\fr v}_{\rm m}$ at the boundary between matter and free space together with the (negative) electric incremental work density $\fr e_{\rm m}$ in the bulk of free space contribute to the incremental action functional as discussed in the sequel.

\subsubsection{Variational Setting}

The \textit{incremental action functional} allows inclusion of a dissipation potential $\fr p_{\rm m}$ and reads
\bea
\!\!\fr A=\fr A(w, \pt w, \gz w)\!\!\!&:=&\!\!\!\int_{\cl T}\bigg[\displaystyle
\int_{\cl B_{\rm m}}[\fr l_{\rm m}(w, \pt w, \gz w, \Grad w, \Grad\pt w, \Grad\gz w, \D_t\pt w, \D_t\gz w)-\fr p_{\rm m}(\pt w/\d t)\d t]\d V\nonumber\\&&\displaystyle\quad\;-\!
\int_{\partial\cl B_{\rm m}}\widehat{\fr v}_{\rm m}(w, \pt w, \gz w)\d A\\\nonumber&&\displaystyle\quad\;-\!
\int_{\cl S_{\rm m}}\fr e_{\rm m}(\Grad w, \Grad\gz w)\d V\bigg]\d t.
\eea

Then, the incremental Hamilton principle requires stationarity of the incremental action functional upon admissible, i.e.\ incremental space-time boundary conditions satisfying material variation $\D_\delta$ of the incremental solution fields $w, \pt w, \gz w$ as
\be
\fr A(w, \pt w, \gz w)\to\mbox{stationary point}.
\ee

Concretely, the stationarity condition for the incremental action functional expands as
\bea{}
\D_\delta\fr A&=&\int_{\cl T}\bigg[\displaystyle
\int_{\cl B_{\rm m}}[\D_\delta\fr l_{\rm m}(w, \pt w, \gz w, \Grad w, \Grad\pt w, \Grad\gz w,\D_t\pt w,\D_t\gz w)-\D_\delta\fr p_{\rm m}(\pt w/\d t)\d t]\d V\nonumber\\ &&\displaystyle\quad\;-\!
\int_{\partial\cl B_{\rm m}}\D_\delta\widehat{\fr v}_{\rm m}(w, \pt w, \gz w)\d A\\
\nonumber &&\displaystyle\quad\;-\!
\int_{\cl S_{\rm m}}\D_\delta\fr e_{\rm m}(\Grad w, \Grad\gz w)\d V\bigg]\d t\doteq 0\quad\forall\D_\delta w, \D_\delta\pt w, \D_\delta\gz w.
\eea

Requiring stationarity of the incremental action functional for arbitrary admissible $\D_\delta w, \D_\delta\pt w, \D_\delta\gz w$ and using the explicit expressions for the incremental action densities as introduced in the above results eventually in the Euler-Lagrange or rather balance equations\footnote{
Based on the dependency of the incremental action functional on the incremental solution fields and their material space-time gradients, variational calculus results in the following Euler-Lagrange equations:\\

$\bullet$ Euler-Lagrange equations in the bulk of matter
\ben
\Div\el D=q^{\rm f}_{\rm m},\qquad
\Div\pt P+\pt b_{\rm m}=\pt s_{\rm m}+\D_t\pt p_{\rm m},\qquad
\Div\gz P+\gz b_{\rm m}=              \D_t\gz p_{\rm m}.
\een

$\bullet$ Euler-Lagrange equations in the bulk of free space
\ben
\Div\el D=    0,\qquad
\Div\gz P=\tz 0.
\een

$\bullet$ Euler-Lagrange equations at the boundary between matter and free space
\ben
\jmp{\el D}\cdot\gz N=\widehat q_{\rm m}^{\rm f},\qquad
     \pt P \cdot\gz N=     \pt t_{\rm m},\qquad
\jmp{\gz P}\cdot\gz N=-    \gz t_{\rm m}.
\een
}
\begin{framed}
\begin{subequations}
\bea
\Div\el D-q^{\rm f}_{\rm m}      \,=   0\;\:\:\;\;\,\qquad\quad\quad\mbox{in}\:\!\quad\cl B_{\rm m}&\mbox{and}&\:\:\:\jmp{\el D}  \cdot\gz N= \widehat q^{\rm f}_{\rm m}\!\!\,\quad\mbox{at}\quad\partial\cl B_{\rm m}\\
\Div\el D\hspace{9.7mm}\,=   0\;\:\:\;\;\,\qquad\quad\quad\mbox{in}\:\!\quad\cl S_{\rm m}&          &\\[2mm]
\Div\pt P\!\;+\pt b_{\rm m}    \!\:=\pt s_{\rm m}+\D_t\pt p_{\rm m}\quad\mbox{in}\quad\cl B_{\rm m}&\mbox{and}&\hspace{5mm}\,\pt P\:\cdot\gz N=\pt t_{\rm m}\;\!\!\!\quad\mbox{at}\quad\partial\cl B_{\rm m}\\[2mm]
\Div\gz P+\gz b_{\rm m}      \,=\,\,\qquad\D_t\gz p_{\rm m}\quad\mbox{in}\quad\cl B_{\rm m}&\mbox{and}&-\jmp{\gz P}  \cdot\gz N=\gz t_{\rm m}\!\quad\mbox{at}\quad\partial\cl B_{\rm m}\\
\Div\gz P\hspace{9.9mm}\,=\tz   0\;\qquad\qquad\quad\mbox{in}\quad\cl S_{\rm m}&          &
\eea
\end{subequations}
\end{framed}

Expanding the balance equation related to the electronic order parameter(s) (micro deformation) in the bulk of matter and neglecting external electronic sources, eventually allows term-by-term comparison with the formulation outlined in \cite{bin2015unified}
\be
\Div      \frac{\partial w^\bullet_{\rm m}}{\partial\pt F}                +J\omega_0\el e=
\frac{\partial w^\circ_{\rm m}}{\partial\pt y}+J\gamma_0\D_t\pt y+\varrho_{\rm m}\D_{tt}\pt y\lto
\div\bigg(\frac{\partial w^\bullet_{\rm m}}{\partial\pt F}\cdot\gz k\bigg)+ \omega_0\el e=
\frac{\partial w^\circ_{\rm s}}{\partial\pt y}+ \gamma_0\D_t\pt y+\varrho_{\rm s}\D_{tt}\pt y.
\ee
The first and second terms left and right are as in the quasi-static situation, for the dynamic situation the third term on the right captures in addition the inertia of the electronic order parameter(s).\\

For completeness, Piola transformation of the balance equations then renders the equivalent expressions in terms of momentum, flux and source densities per unit area and volume, respectively, in the spatial configuration
\begin{subequations}
\bea
\div\el d-q^{\rm f}_{\rm s}      \,=   0\;\:\:\:\;\;\,\qquad\quad\quad\mbox{in}\:\!\quad\cl B_{\rm s}&\mbox{and}&\:\:\:\jmp{\el d}  \cdot\gz n= \widehat q^{\rm f}_{\rm s}\!\!\,\quad\mbox{at}\quad\partial\cl B_{\rm s}\\
\div\el d\hspace{8.8mm}\,=   0\;\;\;\;\;\;\qquad\quad\quad\mbox{in}\:\!\quad\cl S_{\rm s}&          &\\[2mm]
\div\pt s\!\;+\pt b_{\rm s}    \:=\pt s_{\rm s}+j\D_t\pt p_{\rm m}\quad\mbox{in}\quad\cl B_{\rm s}&\mbox{and}&\hspace{5mm}\,\pt s\:\cdot\gz n=\pt t_{\rm s}\quad\;\!\!\!\mbox{at}\quad\partial\cl B_{\rm s}\\[2mm]
\div\gz s\,+\:\!\gz b_{\rm s}     \;=\qquad j\D_t\gz p_{\rm m}\quad\mbox{in}\quad\cl B_{\rm s}&\mbox{and}&-\jmp{\gz s}  \cdot\gz n=\gz t_{\rm s}\!\!\:\quad\mbox{at}\quad\partial\cl B_{\rm s}\\
\div\gz s\hspace{9.7mm}\,=\tz   0\;\,\qquad\qquad\quad\mbox{in}\quad\cl S_{\rm s}&          &
\eea
\end{subequations}

Note the total electronic internal source
density $\pt s_{\rm s}:=\pt s_{\rm s}^\sharp+\pt s_{\rm s}^\flat$ appearing in the balance equation related to the electronic order parameter(s).\\

This concludes derivation of the balance equations for the electric, electronic and mechanical sub-problems, embracing energetic as well as dissipative cases, from Hamilton's principle.

\section{Hamilton Equations}

Hamilton's equations are alternative to Hamilton's principle of least action when
describing the dynamics of a system, however in terms of the evolution of the phases space coordinates rather than the state space coordinates. Thereby, the Hamiltonian follows from a Legendre transformation of the Lagrangian in order to exchange the velocities in state space by their corresponding momenta in phase space. The Hamiltonian setting proves beneficial for dynamical systems with symmetries, i.e.\ when certain momenta are conserved. In the sequel, we will derive Hamilton's equations and thus the ensuing balance equations for energetic and, as a novelty per se, dissipative cases when modelling light-matter interaction in photo-active polymers.

\subsection{Energetic Case}

For the energetic case we first expand on the total Hamiltonian energy density before examining the pertinent variational setting.

\subsubsection{Hamiltonian Energy}

Formally, the \textit{total Hamiltonian energy density} $h^\lambda_{\rm m}=J h^\lambda_{\rm s}$ in the bulk of matter (for the notation see below) follows from Legendre transformation of the total Lagrangian energy density $l_{\rm m}$ exchanging the velocities $v,\pt v, \gz v$ for the corresponding momenta $p_{\rm m}, \pt p_{\rm m}, \gz p_{\rm m}$, i.e.\ by seeking for the supremum
\be
\sup_{v,\pt v, {\scriptsize\gz v}}\{p_{\rm m}\,v+\pt p_{\rm m}\cdot\pt v+\gz p_{\rm m}\cdot\gz v-
l_{\rm m}(y, \pt y, \gz y, \el E, \pt F, \gz F, \pt v, \gz v)\}.
\ee

The total Lagrangian energy density $l_{\rm m}$ does however not depend on the velocity $v$ of the electric potential $y$, thus according to the Dirac theory \cite{dirac1950generalized} $l_{\rm m}$ qualifies as \textit{degenerate}. Consequently, the momentum $p_{\rm m}$ conjugate to $v$ satisfies the
constraint
\be
p_{\rm m}\doteq 0
\ee

that requires enforcement via an additional Lagrange multiplier $\lambda$. It should be noted that, while the electric potential may vary over time, there is no resulting impuls as no corrolated mass exists.\\

Thus, the explicit representation of the total Hamiltonian energy density $h^\lambda_{\rm m}$ in the bulk of matter (whereby the notation $h^\lambda_{\rm m}$ shall indicate inclusion of the Lagrange multiplier $\lambda$) reads

\be
h^\lambda_{\rm m}(y, \pt y, \gz y, \el E, \pt F, \gz F, p_{\rm m}, \pt p_{\rm m}, \gz p_{\rm m}, \lambda)=
t_{\rm m}(\pt p_{\rm m}, \gz p_{\rm m})+
u_{\rm m}(y, \pt y, \gz y, \el E, \pt F, \gz F)+\lambda\, p_{\rm m}.
\ee

Here, the \textit{total dual kinetic energy density} $t_{\rm m}=J t_{\rm s}$ in the bulk of matter that is parameterized in the momenta $\pt p_{\rm m}$ and $\gz p_{\rm m}$ rather than in the velocities $\pt v$ and $\gz v$, respectively, follows likewise from Legendre transformation and reads explicitly
\be
t_{\rm m}=t_{\rm m}(\pt p_{\rm m},\gz p_{\rm m}):=\sup_{\pt v, {\scriptsize\gz v}}\{\pt p_{\rm m}\cdot\pt v+\gz p_{\rm m}\cdot\gz v-k_{\rm m}(\pt v,\gz v)\}=
\frac{1}{2}\frac{1}{\varrho_{\rm m}}\pt p_{\rm m}\cdot\pt p_{\rm m}+\frac{1}{2}\frac{1}{\rho_{\rm m}}\gz p_{\rm m}\cdot\gz p_{\rm m}.
\ee


Finally, the total Hamiltonian energy density $e^\lambda_{\rm m}=J e^\lambda_{\rm s}$ in the bulk of free space follows from Legendre transformation of the electric energy density $e_{\rm m}$ exchanging the velocities $v,\gz v$ (see below) for the corresponding momenta $p_{\rm m}, \gz p_{\rm m}$, i.e.\ by seeking for the supremum
\be
\sup_{v, {\scriptsize\gz v}}\{p_{\rm m}\,v+\gz p_{\rm m}\cdot\gz v+e_{\rm m}(\el E, \gz F)\}.
\ee

The electric energy density $e_{\rm m}$ does, however, not depend on the velocities $v$ and $\gz v$ of the electric potential $y$ and the deformation map $\gz y$, respectively. Thus, in line with the Dirac theory for degenerate Lagrangians, the momenta $p_{\rm m}$ and $\gz p_{\rm m}$ conjugate to $v$ and $\gz v$, respectively, satisfy the constraints
\be
p_{\rm m}\doteq 0\und\gz p_{\rm m}\doteq\tz 0
\ee
that require enforcement via additional Lagrange multipliers $\lambda$ and $\gz\lambda$, respectively.\\

Thus, the explicit representation of the total Hamiltonian energy density $e^\lambda_{\rm m}$ in the bulk of free space (whereby the notation $ e^\lambda_{\rm m}$ shall indicate inclusion of the Lagrange multipliers $\lambda$ and $\gz\lambda$) reads
\be
e^\lambda_{\rm m}(\el E, \gz F, p_{\rm m}, \gz p_{\rm m}, \lambda, \gz\lambda)=
e_{\rm m}(\el E, \gz F)+\lambda\, p_{\rm m}+\gz\lambda\cdot\gz p_{\rm m}.
\ee

The total Hamiltonian energy densities $h^\lambda_{\rm m}$ and $e^\lambda_{\rm m}$ in the bulk of matter and free space, respectively, as well as the total external potential energy density $\widehat v_{\rm m}$ at the boundary between matter and free space contribute to the Hamiltonian energy functional as discussed in the sequel.

\subsubsection{Variational Setting}

To begin with, we define the Hamiltonian energy functional eventually rendering Hamilton's equations as
\bea
H=H(y, \pt y, \gz y, p_{\rm m}, \pt p_{\rm m}, \gz p_{\rm m}, \lambda, \gz\lambda)&:=&\displaystyle
\int_{\cl B_{\rm m}}h^\lambda_{\rm m}(y, \pt y, \gz y, \el E, \pt F, \gz F, p_{\rm m}, \pt p_{\rm m}, \gz p_{\rm m}, \lambda)\d V\nonumber\\&+&
\int_{\partial\cl B_{\rm m}}\widehat v_{\rm m}(y, \pt y, \gz y)\d A\\\nonumber&+&
\int_{\cl S_{\rm m}}e^\lambda_{\rm m}(\el E, \gz F, p_{\rm m}, \gz p_{\rm m}, \lambda, \gz\lambda)\d V.
\eea

Then, with admissible material variations $\D_\delta$ of the phase space $y, \pt y, \gz y, p_{\rm m}, \pt p_{\rm m}, \gz p_{\rm m}$, Hamilton's equations result from requiring
\be
{}\D_{\{t\delta\}}\left[
\int_{\cl D_{\rm m}}\![y\,p_{\rm m}+\gz y\cdot\gz p_{\rm m}]\d V\!+\!
\int_{\cl B_{\rm m}}\!\pt y\cdot\pt p_{\rm m}\d V\right]
\doteq\D_\delta H\quad\forall
\D_\delta y, \D_\delta\pt y, \D_\delta\gz y, \D_\delta p_{\rm m}, \D_\delta\pt p_{\rm m}, \D_\delta\gz p_{\rm m},
\ee
whereby $\cl D_{\rm m}:=\cl B_{\rm m}\cup\cl S_{\rm m}$ denotes the entire solution domain and $\D_{\{t\delta\}}(\bullet, \circ)$ defines the Poisson-bracket-type combination $\D_t(\bullet)\D_\delta(\circ)-\D_\delta(\bullet)\D_t(\circ)$ of material time derivatives and variations\footnote{Interestingly, in terms of the symplectic matrix, $\D_{\{t\delta\}}(\bullet, \circ):=\D_t(\bullet)\D_\delta(\circ)-\D_\delta(\bullet)\D_t(\circ)$ expresses as
\ben
\D_{\{t\delta\}}(\bullet, \circ):=
[\D_t(\bullet), \D_t(\circ)]
\left[\begin{array}{cc}
\phantom{-}0&1\\
         - 1&0
\end{array}\right]
\left[\begin{array}{c}
\D_\delta(\bullet)\\
\D_\delta(\circ)
\end{array}\right]=-
[\D_\delta(\bullet), \D_\delta(\circ)]
\left[\begin{array}{cc}
\phantom{-}0&1\\
         - 1&0
\end{array}\right]
\left[\begin{array}{c}
\D_t(\bullet)\\
\D_t(\circ)
\end{array}\right].
\een
Furthermore, the variation $\D_\delta H$ of a generic Hamiltonian function $H=H(\bullet,\circ)$ reads as
\ben
\D_\delta H=
\left[\frac{\partial H}{\partial (\bullet)}, \frac{\partial H}{\partial (\circ)}\right]
\left[\begin{array}{c}
\D_\delta(\bullet)\\[2mm]
\D_\delta(\circ)
\end{array}\right]=
[\D_\delta(\bullet),\D_\delta(\circ)]
\left[\begin{array}{c}
\displaystyle\frac{\partial H}{\partial (\bullet)}\\[3mm]
\displaystyle\frac{\partial H}{\partial (\circ)}
\end{array}\right].
\een
Thus, finally, due to the skew-symmetry of the symplectic matrix and the arbitrariness of the admissible variations, Hamilton's equations eventually result as
\ben
\left[\begin{array}{c}
\D_t(\bullet)\\
\D_t(\circ)
\end{array}\right]=
\left[\begin{array}{cc}
\phantom{-}0&1\\
         - 1&0
\end{array}\right]
\left[\begin{array}{c}
\displaystyle\frac{\partial H}{\partial (\bullet)}\\[3mm]
\displaystyle\frac{\partial H}{\partial (\circ)}
\end{array}\right]\with
\left[\begin{array}{cc}
\phantom{-}0&1\\
         - 1&0
\end{array}\right]
\left[\begin{array}{cc}
\phantom{-}0&1\\
         - 1&0
\end{array}\right]=-
\left[\begin{array}{cc}
1&0\\
0&1
\end{array}\right].
\een
In case of a Hamiltonian functional, variational derivatives substitute the partial derivatives of $H$.
}.\\

Concretely, the material variation of the Hamiltonian functional expands as
\bea{}
\D_\delta H&=&\displaystyle
\int_{\cl B_{\rm m}}\D_\delta h^\lambda_{\rm m}(y, \pt y, \gz y, \el E, \pt F, \gz F, p_{\rm m}, \pt p_{\rm m}, \gz p_{\rm m}, \lambda)\d V
\nonumber\\&+&\displaystyle
\int_{\partial\cl B_{\rm m}}\D_\delta\widehat v_{\rm m}(y, \pt y, \gz y)\d A
\\\nonumber&+&\displaystyle
\int_{\cl S_{\rm m}}\D_\delta e^\lambda_{\rm m}(\el E, \gz F, p_{\rm m}, \gz p_{\rm m}, \lambda, \gz\lambda)\d V.
\eea

Evaluating the above (Hamiltonian) requirement for arbitrary admissible $\D_\delta y, \D_\delta\pt y, \D_\delta\gz y, \D_\delta p_{\rm m},$ $\D_\delta\pt p_{\rm m}, \D_\delta\gz p_{\rm m}$ and using the constitutive relations as introduced in the above,
Hamilton's equations result in the following balance equations\footnote{Based on the dependency of the Hamiltonian energy functional on the phase space coordinates and their material space gradients, Hamilton's equations, when using the
abbreviations  $p:= p_{\rm m}$, $\pt p:=\pt p_{\rm m}$ and $\gz p:=\gz p_{\rm m}$, read as:\\

$\bullet$ Hamilton equations in the bulk of matter
\ben
\left[\begin{array}{c}
\D_t y\\
\D_t p
\end{array}\right]=
\left[\begin{array}{cc}
\phantom{-}O&I\\
         - I&O
\end{array}\right]
\left[\begin{array}{c}
\displaystyle\frac{\delta   h^\lambda_{\rm m}}{\delta   y}\\[2mm]
\displaystyle\frac{\partial h^\lambda_{\rm m}}{\partial p}
\end{array}\right],\quad
\left[\begin{array}{c}
\D_t\pt y\\
\D_t\pt p
\end{array}\right]=
\left[\begin{array}{cc}
\phantom{-}\pt O&\pt I\\
-\pt I&\pt O
\end{array}\right]
\left[\begin{array}{c}
\displaystyle\frac{\delta   h^\lambda_{\rm m}}{\delta   \pt y}\\[2mm]
\displaystyle\frac{\partial h^\lambda_{\rm m}}{\partial \pt p}
\end{array}\right],\quad
\left[\begin{array}{c}
\D_t\gz y\\
\D_t\gz p
\end{array}\right]=
\left[\begin{array}{cc}
\phantom{-}\gz O&\gz I\\
-\gz I&\gz O
\end{array}\right]
\left[\begin{array}{c}
\displaystyle\frac{\delta   h^\lambda_{\rm m}}{\delta   \gz y}\\[2mm]
\displaystyle\frac{\partial h^\lambda_{\rm m}}{\partial \gz p}
\end{array}\right].
\een

$\bullet$ Hamilton equations in the bulk of free space
\ben
\left[\begin{array}{c}
\D_t y\\
\D_t p
\end{array}\right]=
\left[\begin{array}{cc}
\phantom{-}O&I\\
         - I&O
\end{array}\right]
\left[\begin{array}{c}
\displaystyle\frac{\delta   e^\lambda_{\rm m}}{\delta   y}\\[2mm]
\displaystyle\frac{\partial e^\lambda_{\rm m}}{\partial p}
\end{array}\right],\quad
%
%
\left[\begin{array}{c}
\D_t\gz y\\
\D_t\gz p
\end{array}\right]=
\left[\begin{array}{cc}
\phantom{-}\gz O&\gz I\\
-\gz I&\gz O
\end{array}\right]
\left[\begin{array}{c}
\displaystyle\frac{\delta   e^\lambda_{\rm m}}{\delta   \gz y}\\[2mm]
\displaystyle\frac{\partial e^\lambda_{\rm m}}{\partial \gz p}
\end{array}\right].
\een

$\bullet$ Hamilton equations at the boundary between matter and free space
\ben
-\left[\frac{\partial e^\lambda_{\rm m}}{\partial\Grad    y}-
       \frac{\partial h^\lambda_{\rm m}}{\partial\Grad    y}\right]\cdot\gz N=-\frac{\partial\widehat v_{\rm m}}{\partial    y},\qquad
       \frac{\partial h^\lambda_{\rm m}}{\partial\Grad\pt y}       \cdot\gz N=-\frac{\partial\widehat v_{\rm m}}{\partial\pt y},\qquad
-\left[\frac{\partial e^\lambda_{\rm m}}{\partial\Grad\gz y}-
       \frac{\partial h^\lambda_{\rm m}}{\partial\Grad\gz y}\right]\cdot\gz N=-\frac{\partial\widehat v_{\rm m}}{\partial\gz y}
\een
}
\begin{framed}
\begin{subequations}
\bea
\lambda\,=\,\,\qquad\D_t y\;\;\;\quad\mbox{in}\quad\cl B_{\rm m}\:\!&&\\
\Div\el D-q^{\rm f}_{\rm m}      \,=   0\;\:\:\;\;\,\qquad\quad\quad\mbox{in}\:\!\quad\cl B_{\rm m}&\mbox{and}&\:\:\:\jmp{\el D}  \cdot\gz N= \widehat q^{\rm f}_{\rm m}\!\!\,\quad\mbox{at}\quad\partial\cl B_{\rm m}\\[1mm]
\lambda\,=\,\,\qquad\D_t y\;\;\;\quad\mbox{in}\quad\cl S_{\rm m}\:\!&&\\
\Div\el D\hspace{9.7mm}\,=   0\;\:\:\;\;\,\qquad\quad\quad\mbox{in}\:\!\quad\cl S_{\rm m}&          &\\[2mm]
\pt p_{\rm m}/\varrho_{\rm m}\,=\:\,\qquad\D_t\pt y\:\:\:\quad\mbox{in}\quad\cl B_{\rm m}&&\\
\Div\pt P\!\;+\pt b_{\rm m}    \!\:=\pt s_{\rm m}^\sharp+\D_t\pt p_{\rm m}\quad\mbox{in}\quad\cl B_{\rm m}&\mbox{and}&\hspace{5mm}\,\pt P\:\cdot\gz N=\pt t_0\quad\mbox{at}\quad\partial\cl B_{\rm m}\\[2mm]
\gz p_{\rm m}/\rho_{\rm m}\,=\;\qquad\D_t\gz y\:\:\:\quad\mbox{in}\quad\cl B_{\rm m}&&\\
\Div\gz P+\gz b_{\rm m}      \,=\,\,\qquad\D_t\gz p_{\rm m}\quad\mbox{in}\quad\cl B_{\rm m}&\mbox{and}&-\jmp{\gz P}  \cdot\gz N=\gz t_0\:\!\quad\mbox{at}\quad\partial\cl B_{\rm m}\\[1mm]
\gz\lambda\,=\,\:\qquad\D_t\gz y\,\:\:\quad\mbox{in}\quad\cl S_{\rm m}&\\
\Div\gz P\hspace{9.9mm}\,=\tz   0\;\qquad\qquad\quad\mbox{in}\quad\cl S_{\rm m}&          &
\eea
\end{subequations}
\end{framed}

For completeness, Piola transformation then renders the equivalent expressions in terms of momentum, flux and source densities per unit area and volume, respectively, in the spatial configuration

\begin{subequations}
\bea
j\lambda\,=\:\,\qquad j\D_t y\;\;\quad\mbox{in}\;\!\quad\cl B_{\rm s}&&\\
\div\el d-q^{\rm f}_{\rm s}      \,=   0\;\:\:\:\;\;\,\qquad\quad\quad\mbox{in}\;\!\quad\cl B_{\rm s}&\mbox{and}&\:\:\:\jmp{\el d}  \cdot\gz n= \widehat q^{\rm f}_{\rm s}\!\!\,\quad\mbox{at}\quad\partial\cl B_{\rm s}\\[1mm]
j\lambda\,=\:\,\qquad j\D_t y\;\;\quad\mbox{in}\;\!\quad\cl S_{\rm s}&&\\
\div\el d\hspace{8.8mm}\,=   0\;\;\;\;\;\;\qquad\quad\quad\mbox{in}\;\!\quad\cl S_{\rm s}&          &\\[2mm]
j\pt p_{\rm s}/\varrho_{\rm s}\,=\,\qquad j\D_t\pt y\:\:\:\quad\mbox{in}\quad\cl B_{\rm s}&&\\
\div\pt s\!\;+\pt b_{\rm s}    \:=\,\pt s_{\rm s}^\sharp+j\D_t\pt p_{\rm m}\quad\mbox{in}\quad\cl B_{\rm s}&\mbox{and}&\hspace{5mm}\,\pt s\:\cdot\gz n=\pt t_{\rm s}\quad\;\!\!\!\mbox{at}\quad\partial\cl B_{\rm s}\\[2mm]
j\gz p_{\rm s}/\rho_{\rm s}\,=\,\qquad j\D_t\gz y\:\:\:\quad\mbox{in}\quad\cl B_{\rm s}&&\\
\div\gz s\,+\:\!\gz b_{\rm s}     \;=\,\qquad j\D_t\gz p_{\rm m}\quad\mbox{in}\quad\cl B_{\rm s}&\mbox{and}&-\jmp{\gz s}  \cdot\gz n=\gz t_{\rm s}\!\!\:\quad\mbox{at}\quad\partial\cl B_{\rm s}\\[1mm]
j\gz\lambda\,=\,\:\!\qquad j\D_t\gz y\;\;\quad\mbox{in}\quad\cl S_{\rm s}&&\\
\div\gz s\hspace{9.7mm}\,=\tz   0\;\;\qquad\qquad\quad\mbox{in}\quad\cl S_{\rm s}&          &
\eea
\end{subequations}

Note the extra equations relating either the momenta or the Lagrange multipliers (that enforce the constraints $p_{\rm m}=0$ in $\cl D_{\rm m}$ and $\gz p_{\rm m}=\tz 0$ in $\cl S_{\rm m}$) to their conjugate velocities.

\subsection{Dissipative Case}

For the dissipative case we consider the incremental total energy density as basis ingredient for the pertinent variational setting that allows inclusion of a dissipation potential.

\subsubsection{Incremental Total Energy}

For an extension towards the dissipative case, we first introduce and abbreviate the increments of the electric, electronic and mechanical momenta as
\be
q:=\d p_{\rm m},\qquad\pt q:=\d\pt p_{\rm m},\qquad\gz q:=\d\gz p_{\rm m}.
\ee

Next, the \textit{incremental 'inertial work density'}, instrumental for Legendre transforming the Lagrangian into the Hamiltonian, follows as the increment of twice the total kinetic energy density
\be
\d[p_{\rm m}\,v+\pt p_{\rm m}\cdot\pt v+\gz p_{\rm m}\cdot\gz v]=
p_{\rm m}\D_t w+\pt p_{\rm m}\cdot\D_t\pt w+\gz p_{\rm m}\cdot\D_t\gz w+v\,q+\pt v\cdot\pt q+\gz v\cdot\gz q.
\ee

Then, the \textit{incremental total energy density} $\fr h^\lambda_{\rm m}=J \fr h^\lambda_{\rm s}$ in the bulk of matter follows formally from Legendre transformation of the incremental action density $\fr l_{\rm m}$ exchanging the incremental velocities $\D_t w, \D_t\pt w, \D_t\gz w$ for the corresponding incremental momenta $q, \pt q, \gz q$, thus resulting in\footnote{
Alternatively with the constitutive relation $\pt p_{\rm m}/\varrho_{\rm m}=:\pt v$ and $\gz p_{\rm m}/\rho_{\rm m}=:\gz v$, the incremental total energy density follows from
\ben
\fr h^\lambda_{\rm m}(w, \pt w, \gz w, \Grad w, \Grad\pt w, \Grad\gz w, q, \pt q, \gz q,\lambda):=\big[\d\pw h^\lambda_{\rm m}(y, \pt y, \gz y, \el E, \pt F, \gz F, p_{\rm m}, \pt p_{\rm m}, \gz p_{\rm m}, \lambda)\,\big]_{\rm fixed\: phase\: space}.
\een
Here, the constraint $p_{\rm m}\doteq 0$ eliminates the incremental Lagrange multiplier $\d\lambda$.
}
\bea
&\fr h^\lambda_{\rm m}=\fr h_{\rm m}(w, \pt w, \gz w, \Grad w, \Grad\pt w, \Grad\gz w, q, \pt q, \gz q, \lambda):=&\\\nonumber&
\d[p_{\rm m}\,v+\pt p_{\rm m}\cdot\pt v+\gz p_{\rm m}\cdot\gz v]-
\fr l_{\rm m}(w, \pt w, \gz w, \Grad w, \Grad\pt w, \Grad\gz w, \D_t\pt w, \D_t\gz w).&
\eea

Consequently, and explicitly incorporating the constraint $p_{\rm m}\doteq 0$ and the explicit result for the Lagrange multiplier $v=\lambda$ as well as the explicit representation of the incremental action density $\fr l_{\rm m}$ in the above, the explicit representation of the incremental total energy density in the bulk of matter follows as
\bea{}
&\fr h^\lambda_{\rm m}(w, \pt w, \gz w, \Grad w, \Grad\pt w, \Grad\gz w, q, \pt q, \gz q,\lambda)=&\\\nonumber
&\displaystyle[\el D\cdot\Grad w+q^{\rm f}_{\rm m}\, w+\lambda\, q]+[\pt P:\Grad\pt w-[\pt b_{\rm m}-\pt s_{\rm m}^\sharp]\cdot\pt w+\pt v\cdot\pt q]+[\gz P:\Grad\gz w-\gz b_{\rm m}\cdot\gz w+\gz v\cdot\gz q].&
\eea

In an entirely analogous fashion, the \textit{incremental total energy density} $\fr e^\lambda_{\rm m}=J \fr e^\lambda_{\rm s}$ in the bulk of free space computes as
\bea{}
&\fr e^\lambda_{\rm m}(\Grad w, \Grad\gz w, q, \gz q,\lambda,\gz\lambda):=&\\\nonumber
&\big[\d e^\lambda_{\rm m}(\el E, \gz F, p_{\rm m}, \gz p_{\rm m}),\lambda,\gz\lambda\,\big]_{\rm fixed\: phase\: space}=
[\el D\cdot\Grad w+\lambda\, q]+[\gz P:\Grad\gz w+\gz\lambda\cdot\gz q].&
\eea

The incremental total energy densities $\fr h^\lambda_{\rm m}$ and  $\fr e^\lambda_{\rm m}$ in the bulk of matter and free space, respectively, as well as the external incremental work density $\widehat{\fr v}_{\rm m}$ at the boundary between matter and free space contribute to the incremental total energy functional as discussed in the sequel.

\subsubsection{Variational Setting}

The \textit{incremental total energy functional} allows inclusion of the dissipation potential $\fr p_{\rm m}$ and reads
\bea
\fr H=\fr H(w, \pt w, \gz w, q, \pt q, \gz q,\lambda,\gz\lambda)&:=&
\int_{\cl B_{\rm m}}[\fr h^\lambda_{\rm m}(w, \pt w, \gz w, \Grad w, \Grad\pt w, \Grad\gz w, q, \pt q, \gz q,\lambda)+\fr p_{\rm m}(\pt w/\d t)\d t]\d V\nonumber\\&+&\displaystyle
\int_{\partial\cl B_{\rm m}}\widehat{\fr v}_{\rm m}(w, \pt w, \gz w)\d A\\\nonumber&+&\displaystyle
\int_{\cl S_{\rm m}}\fr e^\lambda_{\rm m}(\Grad w, \Grad\gz w, q, \gz q,\lambda,\gz\lambda)\d V.
\eea

Then, with admissible material variations $\D_\delta$ of the incremental phase space $w, \pt w, \gz w, q, \pt q, \gz q$, Hamilton's equations result from requiring incrementally
\footnote{
The increments of the integrands on the left-hand side expand as
\ben
\d[y\, p_{\rm m}+\gz y\cdot\gz p_{\rm m}]=
w\,p_{\rm m}+\gz w\cdot\gz p_{\rm m}+y\,q+\gz y\cdot\gz q\und
\d[\pt y\cdot\pt p_{\rm m}]=\pt w\cdot\pt p_{\rm m}+\pt y\cdot\pt q.
\een
Thus, application of $\D_{\{t\delta\}}$ for material time derivatives of the incremental phase space $w, \pt w, \gz w, q, \pt q, \gz q$ \textit{and} the phase space $y, \pt y, \gz y, p_{\rm m}, \pt p_{\rm m}, \gz p_{\rm m}$ as well as for material variations $\D_\delta w, \D_\delta\pt w, \D_\delta\gz w, \D_\delta q, \D_\delta\pt q, \D_\delta\gz q$ of \textit{only} the incremental phase space  (i.e.\ with \textit{vanishing} material variations $\D_\delta y, \D_\delta\pt y, \D_\delta\gz y, \D_\delta p_{\rm m}, \D_\delta\pt p_{\rm m}, \D_\delta\gz p_{\rm m}$), renders explicitly
\ben
\D_{\{t\delta\}}\d[y\, p_{\rm m}+\gz y\cdot\gz p_{\rm m}]=
\D_t y\D_\delta q+\D_t\gz y\cdot\D_\delta\gz q-\D_\delta w\,\D_t p_{\rm m}-\D_\delta\gz w\cdot\D_t\gz p_{\rm m}\und
\D_{\{t\delta\}}\d[\pt y\cdot\pt p_{\rm m}]=\D_t\pt y\cdot\D_\delta\pt q-\D_\delta\pt w\cdot\D_t\pt p_{\rm m}.
\een
}
\be
{}\D_{\{t\delta\}}\left[
\int_{\cl D_{\rm m}}\!\!\!\d[y\,p_{\rm m}+\gz y\cdot\gz p_{\rm m}]\d V\!+\!
\int_{\cl B_{\rm m}}\!\!\!\d[\pt y\cdot\pt p_{\rm m}]\d V\right]
\doteq\D_\delta\fr H\quad\forall
\D_\delta w, \D_\delta\pt w, \D_\delta\gz w, \D_\delta q, \D_\delta\pt q, \D_\delta\gz q.
\ee

Concretely, the material variation of the incremental total energy functional expands as
\bea{}
\D_\delta\fr H&=&\displaystyle
\int_{\cl B_{\rm m}}[\D_\delta\fr h^\lambda_{\rm m}(w, \pt w, \gz w, \Grad w, \Grad\pt w, \Grad\gz w, q, \pt q, \gz q, \lambda)+\D_\delta\fr p_{\rm m}(\pt w/\d t)\d t]\d V
\nonumber\\&+&\displaystyle
\int_{\partial\cl B_{\rm m}}\D_\delta\widehat{\fr v}_{\rm m}(w, \pt w, \gz w)\d A
\\\nonumber&+&\displaystyle
\int_{\cl S_{\rm m}}\D_\delta\fr e^\lambda_{\rm m}(\Grad w, \Grad\gz w, q, \gz q, \lambda, \gz\lambda)\d V.
\eea

Evaluating the above incremental (Hamiltonian) requirement for arbitrary admissible $\D_\delta w, \D_\delta\pt w, \D_\delta\gz w$, $\D_\delta q, \D_\delta\pt q, \D_\delta\gz q$ and using the expressions for the incremental total energy densities as introduced in the above,
Hamilton's equations result eventually in the following balance equations\footnote{
Based on the dependency of the incremental total energy functional on the incremental phase space coordinates and their material space gradients, Hamilton's equations read as:\\

$\bullet$ Hamilton equations in the bulk of matter
\ben
\left[\begin{array}{c}
\D_t y\\[2mm]
\D_t p
\end{array}\right]=
\left[\begin{array}{cc}
\phantom{-}O&I\\
         - I&O
\end{array}\right]
\left[\begin{array}{c}
\displaystyle\frac{\delta  \fr h^\lambda_{\rm m}}{\delta   w}\\[2mm]
\displaystyle\frac{\partial\fr h^\lambda_{\rm m}}{\partial q}
\end{array}\right],\quad
\left[\begin{array}{c}
\D_t\pt y\\[2mm]
\D_t\pt p
\end{array}\right]=
\left[\begin{array}{cc}
\phantom{-}\pt O&\pt I\\
-\pt I&\pt O
\end{array}\right]
\left[\begin{array}{c}
\displaystyle\frac{\delta  \fr h^\lambda_{\rm m}}{\delta   \pt w}\\[2mm]
\displaystyle\frac{\partial\fr h^\lambda_{\rm m}}{\partial \pt q}
\end{array}\right],\quad
\left[\begin{array}{c}
\D_t\gz y\\[2mm]
\D_t\gz p
\end{array}\right]=
\left[\begin{array}{cc}
\phantom{-}\gz O&\gz I\\
-\gz I&\gz O
\end{array}\right]
\left[\begin{array}{c}
\displaystyle\frac{\delta  \fr h^\lambda_{\rm m}}{\delta   \gz w}\\[2mm]
\displaystyle\frac{\partial\fr h^\lambda_{\rm m}}{\partial \gz q}
\end{array}\right].
\een

$\bullet$ Hamilton equations in the bulk of free space
\ben
\left[\begin{array}{c}
\D_t y\\[2mm]
\D_t p
\end{array}\right]=
\left[\begin{array}{cc}
\phantom{-}O&I\\
         - I&O
\end{array}\right]
\left[\begin{array}{c}
\displaystyle\frac{\delta  \fr e^\lambda_{\rm m}}{\delta   w}\\[2mm]
\displaystyle\frac{\partial\fr e^\lambda_{\rm m}}{\partial q}
\end{array}\right],\quad
%
%
\left[\begin{array}{c}
\D_t\gz y\\[2mm]
\D_t\gz p
\end{array}\right]=
\left[\begin{array}{cc}
\phantom{-}\gz O&\gz I\\
-\gz I&\gz O
\end{array}\right]
\left[\begin{array}{c}
\displaystyle\frac{\delta  \fr e^\lambda_{\rm m}}{\delta   \gz w}\\[2mm]
\displaystyle\frac{\partial\fr e^\lambda_{\rm m}}{\partial \gz q}
\end{array}\right].
\een

$\bullet$ Hamilton equations at the boundary between matter and free space
\ben
-\left[\frac{\partial\fr e^\lambda_{\rm m}}{\partial\Grad    w}-
       \frac{\partial\fr h^\lambda_{\rm m}}{\partial\Grad    w}\right]\cdot\gz N=-\frac{\partial\widehat{\fr v}_{\rm m}}{\partial    w},\qquad
       \frac{\partial\fr h^\lambda_{\rm m}}{\partial\Grad\pt w}       \cdot\gz N=-\frac{\partial\widehat{\fr v}_{\rm m}}{\partial\pt w},\qquad
-\left[\frac{\partial\fr e^\lambda_{\rm m}}{\partial\Grad\gz w}-
       \frac{\partial\fr h^\lambda_{\rm m}}{\partial\Grad\gz w}\right]\cdot\gz N=-\frac{\partial\widehat{\fr v}_{\rm m}}{\partial\gz w}
\een
}
\begin{framed}
\begin{subequations}
\bea
\lambda\,=\,\,\qquad\D_t y\;\;\;\quad\mbox{in}\quad\cl B_{\rm m}\:\!&\\
\Div\el D-q^{\rm f}_{\rm m}      \,=   0\;\:\:\;\;\,\qquad\quad\quad\mbox{in}\:\!\quad\cl B_{\rm m}&\mbox{and}&\:\:\:\jmp{\el D}  \cdot\gz N= \widehat q^{\rm f}_{\rm m}\!\!\,\quad\mbox{at}\quad\partial\cl B_{\rm m}\\[1mm]
\lambda\,=\,\,\qquad\D_t y\;\;\;\quad\mbox{in}\quad\cl S_{\rm m}\:\!&\\
\Div\el D\hspace{9.7mm}\,=   0\;\:\:\;\;\,\qquad\quad\quad\mbox{in}\:\!\quad\cl S_{\rm m}&          &\\[2mm]
\pt v\,=\:\,\qquad\D_t\pt y\:\:\:\quad\mbox{in}\quad\cl B_{\rm m}&\\
\Div\pt P\!\;+\pt b_{\rm m}    \!\:=\pt s_{\rm m}+\D_t\pt p_{\rm m}\quad\mbox{in}\quad\cl B_{\rm m}&\mbox{and}&\hspace{5mm}\,\pt P\:\cdot\gz N=\pt t_0\quad\mbox{at}\quad\partial\cl B_{\rm m}\\[2mm]
\gz v\,=\;\qquad\D_t\gz y\:\:\:\quad\mbox{in}\quad\cl B_{\rm m}&\\
\Div\gz P+\gz b_{\rm m}      \,=\,\,\qquad\D_t\gz p_{\rm m}\quad\mbox{in}\quad\cl B_{\rm m}&\mbox{and}&-\jmp{\gz P}  \cdot\gz N=\gz t_0\:\!\quad\mbox{at}\quad\partial\cl B_{\rm m}\\[1mm]
\gz\lambda\,=\,\:\qquad\D_t\gz y\,\:\:\quad\mbox{in}\quad\cl S_{\rm m}&\\
\Div\gz P\hspace{9.9mm}\,=\tz   0\;\qquad\qquad\quad\mbox{in}\quad\cl S_{\rm m}&          &
\eea
\end{subequations}
\end{framed}

Note again the total electronic internal source
density $\pt s_{\rm s}:=\pt s_{\rm s}^\sharp+\pt s_{\rm s}^\flat$ appearing in the balance equation related to the electronic order parameter(s).\\

For completeness, Piola transformation then renders the equivalent expressions in terms of momentum, flux and source densities per unit area and volume, respectively, in the spatial configuration

\begin{subequations}
\bea
j\lambda\,=\:\,\qquad j\D_t y\;\;\quad\mbox{in}\;\!\quad\cl B_{\rm s}&&\\
\div\el d-q^{\rm f}_{\rm s}      \,=   0\;\:\:\:\;\;\,\qquad\quad\quad\mbox{in}\;\!\quad\cl B_{\rm s}&\mbox{and}&\:\:\:\jmp{\el d}  \cdot\gz n= \widehat q^{\rm f}_{\rm s}\!\!\,\quad\mbox{at}\quad\partial\cl B_{\rm s}\\[1mm]
j\lambda\,=\:\,\qquad j\D_t y\;\;\quad\mbox{in}\;\!\quad\cl S_{\rm s}&&\\
\div\el d\hspace{8.8mm}\,=   0\;\;\;\;\;\;\qquad\quad\quad\mbox{in}\;\!\quad\cl S_{\rm s}&          &\\[2mm]
j\pt p_{\rm s}/\varrho_{\rm s}\,=\,\qquad j\D_t\pt y\:\:\:\quad\mbox{in}\quad\cl B_{\rm s}&&\\
\div\pt s\!\;+\pt b_{\rm s}    \:=\,\pt s_{\rm s}+j\D_t\pt p_{\rm m}\quad\mbox{in}\quad\cl B_{\rm s}&\mbox{and}&\hspace{5mm}\,\pt s\:\cdot\gz n=\pt t_{\rm s}\quad\;\!\!\!\mbox{at}\quad\partial\cl B_{\rm s}\\[2mm]
j\gz p_{\rm s}/\rho_{\rm s}\,=\,\qquad j\D_t\gz y\:\:\:\quad\mbox{in}\quad\cl B_{\rm s}&&\\
\div\gz s\,+\:\!\gz b_{\rm s}     \;=\,\qquad j\D_t\gz p_{\rm m}\quad\mbox{in}\quad\cl B_{\rm s}&\mbox{and}&-\jmp{\gz s}  \cdot\gz n=\gz t_{\rm s}\!\!\:\quad\mbox{at}\quad\partial\cl B_{\rm s}\\[1mm]
j\gz\lambda\,=\,\:\!\qquad j\D_t\gz y\;\;\quad\mbox{in}\quad\cl S_{\rm s}&&\\
\div\gz s\hspace{9.7mm}\,=\tz   0\;\;\qquad\qquad\quad\mbox{in}\quad\cl S_{\rm s}&          &
\eea
\end{subequations}

This concludes derivation of the balance equations for the electric, electronic and mechanical sub-problems, embracing energetic as well as dissipative cases, from Hamilton's equations.

\section{Conclusion}

Continuum modeling of light-matter interaction in photo-active polymers is instrumental when designing and optimising devices with the technologically attractive capacity for remote and contact-free actuation by light. Current research in organic chemistry focusses on synthesis and characterization of a variety of polymer compounds involving different molecular photo-switches. Thus, there is great promise for future development of photo-mechanically coupled material compositions. These also include options for mechanically soft photo-active polymers, thus asking for a continuum framework providing geometrically exact description of the deformation and related kinematic quantities. Continuum formulations are key pre-requisite for computational simulations of devices based on, e.g., variational (Galerkin-type) approaches such as the finite element method. In this contribution we provide the necessary preliminaries, such as continuum formulations of the solution fields, energy densities and constitutive relations, entering the electric, electronic and mechanical sub-problems, for a comprehensive account on the variational setting of photo-mechanics. Based thereon, we demonstrated how to variationally cast the pertinent equilibrium and balance equations for energetic as well as, noteworthy, dissipative cases. In combination with Hamilton's equations, especially the latter case is a novelty per se. In conclusion, we established a geometrically exact variational continuum framework of light-matter interaction in photo-active polymers that allows for analytical and computational investigations of photo-mechanical devices. Forthcoming contributions will focus on the corresponding continuum thermodynamics and the computational setting, among further extensions.

\vspace{-0.3cm}
\bibliography{Photostriction.bib}
\bibliographystyle{ieeetr}\label{bib}

\end{document}